\documentclass[preprint,12pt]{elsarticle}
\biboptions{sort&compress}

\usepackage{amssymb}
\usepackage{amsmath}
\usepackage{float}
\usepackage{subfig}
\usepackage[export]{adjustbox}
\usepackage[hidelinks]{hyperref}
\usepackage{changepage}
\journal{Energy Policy}

\begin{document}

\begin{frontmatter}



\title{Power sector effects of green hydrogen production in Germany}
% Vorschlag: Power sector effects of green hydrogen production in Germany 2030


\author[DIW]{Dana Kirchem\corref{cor1}}
\ead{dkirchem@diw.de}
\author[DIW]{Wolf-Peter Schill}
\ead{wschill@diw.de}

\affiliation[DIW]{organization={DIW Berlin},%Department and Organization
            addressline={Mohrenstrasse 58}, 
            city={Berlin},
            postcode={10117}, 
            country={Germany}}

\begin{abstract}
The use of green hydrogen can support the decarbonization of sectors which are difficult to electrify, such as industry or heavy transport. Yet, the wider power sector effects of providing green hydrogen are not well understood so far. We use an open-source electricity sector model to investigate potential power sector interactions of three alternative supply chains for green hydrogen in Germany in the year 2030. We distinguish between model settings in which Germany is modeled as an electric island versus embedded in an interconnected system with its neighboring countries, as well as settings with and without technology-specific capacity bounds on wind energy. The findings suggest that large-scale hydrogen storage can provide valuable flexibility to the power system in settings with high renewable energy shares. These benefits are more pronounced in the absence of flexibility from geographical balancing. We further find that the effects of green hydrogen production on the optimal generation portfolio strongly depend on the model assumptions regarding capacity expansion potentials. We also identify a potential distributional effect of green hydrogen production at the expense of other electricity consumers, of which policy makers should be aware.  
\end{abstract}


%    \item Hydrogen storage flexibility in Germany is assessed with a capacity expansion% model.
%    \item Cavern storage of hydrogen provides valuable temporal flexibility. 
%    \item Hydrogen road transport costs can diminish flexibility benefits.
%    \item Flexibility benefits decrease when cheaper flexibility options are available. 


\end{frontmatter}

\newpage
\section*{Nomenclature}
\begingroup
\renewcommand{\arraystretch}{1.25} % Default value: 1
\begin{table}[H]
    \begin{tabular}{ p{4cm} p{8cm}}
    $\eta^{ntc}$ & Factor correcting for transmission losses \newline based on distance [0,1] \\
    $\phi^{RES}$ & Renewable share target value [0,1] \\
    $E^{H2}$ & Electricity required for electrolysis [MWh] \\
    $F$     & Electricity flow between countries [MWh] \\
    $fossil$ & Set of fossil fuel technologies \\
    $G^{fossil}$ & Power generation from fossil fuels [MWh] \\
    $G^{RES}$ & Power generation from renewable energy sources [MWh] \\
    $H$    & Set of hours \\
    $H2$ & Hydrogen gas \\
    $H2_{STO}$ & Hydrogen storage \\
    $H2_{TRANS}$ & Hydrogen transport \\
    $INVCOST$ & Investment and fixed costs [Euro] \\
    $L$ & Set of cross-border transmission lines\\
    $N$     & Set of nodes \\
    $NTC$   & Net transfer capacities \\
    $RES$   & Set of renewable energy sources \\
    $rsvr$  & Hydro-power reservoirs \\
    $STO$   & Set of electricity storage technologies \\
    $TECH$  & Set of power generation technologies \\
    $VARCOST$ & Variable costs [Euro] \\
    $Z$     & System costs [Euro] \\
    \end{tabular}%
  \label{tab:nomenclature}%
\end{table}%
\endgroup

\section{Introduction}\label{sec: introduction}
It is well established by now that hydrogen produced from renewable energy sources (green hydrogen) can support the decarbonization of sectors which are difficult to electrify, such as industry or heavy transport \cite{deconinck_2018,IPCC_2022}. Thus, hydrogen production is expected to massively increase in the upcoming years, with electrolysis as the main process to produce green hydrogen. At the same time, the interactions of green hydrogen with the power sector are not yet well understood. On the one hand, green hydrogen may contribute to renewable integration if production is sufficiently flexible. On the other hand, the increased electricity demand may create additional challenges of renewable integration.\\

In this study, we use an open-source power sector optimization model to investigate the effects of different hydrogen supply chains on the optimal power plant portfolio, its dispatch, costs, and prices. In particular, we are interested in the following questions: How does the flexible production of hydrogen affect renewable energy generation? Does the availability of hydrogen storage reduce the capacity needs and usage of electricity storage options? And to what extent do different hydrogen storage options results in comparative system cost savings? 

To do so, we differentiate three hydrogen supply chains. In the first model setting, hydrogen can be generated and stored in high-pressure tanks which are located on the site of hydrogen demand. In a second setting, we assume that lower-cost cavern storage is located at the site of hydrogen demand instead of higher-cost tank storage. It can be expected that this storage option results in lower overall system costs and higher renewable energy integration, since it is cheaper than high-pressure tank storage and can act as a seasonal storage. However, in most cases, it might not be feasible to locate hydrogen demand close to a suitable cavern location. Thus, electrolysis and cavern storage are assumed to be distant from the location of hydrogen demand in a third model setting, and trailer-based transport of hydrogen is required. The additional transport costs might eat up some of the benefits of cavern storage. We apply the model to a 2030 scenarios of the central European power sector. For hydrogen, we focus on Germany, since the country has ambitious political targets for promoting the domestic use and production of green hydrogen, having recently declared an electrolyzer capacity target of 10 GW by 2030. \footnote{See \url{https://openenergytracker.org/en/docs/germany/hydrogen/} for the monitoring of German electrolysis targets.}\\

\section{Literature}\label{sec:literature}
%Brief review of relevant literature
There is a growing literature on techno-economic analyses of green hydrogen. Many studies focus on the potentials and costs of hydrogen generation, partly with great technological detail, but hardly focus on electricity sector aspects. The quantities or prices of electricity used for electrolysis are often based on exogenous assumptions. For example, Schlund and Theile look into the optimal dispatch of electrolyzers in order to reduce emission intensity with the help of a cost-optimisation model for electrolyzer operations, which takes the relevant electricity market parameter as an input \cite{schlund_2022}. Other studies of hydrogen for mobility purposes in Germany \cite{welder_2018} or Great Britain \cite{samsatli_2016} assume that electrolysis is supplied with electricity from wind power that is built and used exclusively for this purpose. Other analyses of green hydrogen in Germany assume that electrolyzers only use renewable surplus energy \cite{robinius_2017,emonts_2019}. Glenk and Reichelstein investigate the prices above which it would be more profitable for wind turbine operators to produce green hydrogen as compared to feeding electricity to the grid in Germany or Texas \cite{glenk_2019}. Other studies focus on differences in production and distribution chains for green hydrogen, relying on exogenous electricity price assumptions \cite{yang_2007,reuss_2017}. Sens et al.~investigate the costs of hydrogen production and supply across European and MENA countries, assuming electrolyzers are linked to purpose-built PV and wind power plants \cite{sens_2022}.\\

Another strand of the literature covers the electricity sector in more detail. For example, Zhang et al.~use a dispatch model to investigate the benefits of temporally flexible hydrogen infrastructure in the Western U.S.~power system \cite{Zhang_2020}. Focusing on Germany, similar models are used to derive hourly prices for electricity used in hydrogen supply chains \cite{runge_2019} or to analyze the effects of different hydrogen infrastructures on the dispatch of power plant capacities \cite{michalski_2017}. Such models are also applied to explore the spatial aspects of hydrogen production and its interactions with electricity grids. For example, vom Scheidt et al.~focus on the differences between uniform and zonal electricity market pricing on hydrogen infrastructure investments in Germany and find lower hydrogen costs in case of spatially electricity price signals \cite{vomscheidt_2022}. Their study is particularly detailed with respect to the types and geographic locations of hydrogen demand locations, but again uses a pure dispatch model, i.e.,~does not investigate the repercussions of green hydrogen on optimal power plant portfolio, and abstracts from temporal flexibility aspects and hydrogen storage. In a case study of the Netherlands, Weimann et al.~explore the role of hydrogen in a power system with high wind penetration \cite{weimann_2021}. Breder et al.~investigate the impact of a potential market split on optimal investments, dispatch and location of electrolyzers in Germany \cite{breder_2022}.In a complementary paper, Lieberwirth and Hobbie analyze the interactions of green hydrogen with the optimal operation of the German transmission system \cite{lieberwirth_2023}.\\

Using capacity expansion models instead of pure dispatch models allows to also assess the effects of green hydrogen on optimal power plant portfolios. For example, Brown et al.~use the model PyPSA to investigate the effects of different sector coupling options in a future European interconnection with high shares of renewable energy \cite{brown_2018}. They find that power-to-gas options in combination with hydrogen or methane storage can help to balance longer-term variations of renewable generation and demand. Yet in their study, green hydrogen is either reconverted to electricity or used for heating (directly or via methanization), but no direct use for green hydrogen, e.g.,~in industry, is modeled. Durakovic et al.~provide an assessment of future grid expansion needs in Europe due to hydrogen production powered by offshore wind production in the North Sea. Using the capacity expansion model EMPIRE, they find that European power generation capacity increases by 50 percent \cite{durakovic_2023}.  

Focusing on Germany, Stöckl et al.~apply a capacity expansion model to explore the trade-off between energy efficiency and temporal flexibility of four different options for supplying electrolysis-based hydrogen~\cite{stoeckl_2021}. Their analysis focuses on hydrogen provision at filling stations for mobility purposes, and only considers the German power sector in isolation. Another capacity expansion model analysis finds that hydrogen and synthetic natural gas play an important role in long-term emission reduction scenarios for France \cite{shirizadeh_2022}. While this analysis covers numerous technologies, its main focus is not on temporal flexibility of green hydrogen. By assumption, low-cost hydrogen caverns can be built without restrictions, so hydrogen supply is very flexible. Another study by \cite{peterssen_2022} focuses on the relationship between hydrogen imports and domestic hydrogen production based on the maximum installable PV capacity and the import price of hydrogen. Based on different hydrogen supply scenarios, they simulate cost-optimal transformation pathways of the German energy system until 2050 using particle swarm optimization. They find that hydrogen is used for primary energy purposes at low import prices, while it becomes less attractive as a primary energy source at higher import prices, but is still required to provide flexibility to the energy system. Not least, flexible generation and use of green hydrogen also receives increasing interest from integrated assessment modelers. For example, it is included in a recent study that explores global low-emission scenarios with very high use of variable renewable energy sources and different electrification options \cite{luderer_2022}. Yet, due to the complexity of the large-scale model REMIND used here, hydrogen storage and its temporal interaction with renewable variability are not modeled in detail.\\

%Own contribution
Summing up, various studies have analyzed the current or future costs and potentials for green hydrogen production. Yet, its interactions with the electricity sector have hardly been studied in detail, particularly concerning the temporal flexibility of different options for hydrogen generation and storage. Our research adds to the previous literature by analyzing the interactions of green hydrogen production and the electricity sector in more detail, including the repercussions of green hydrogen production on optimal power plant and storage portfolios. In doing so, we focus on temporal flexibility aspects and endogenously model not only electrolyzer capacity, but also the capacities of two different types of on- or off-site hydrogen storage. To ensure transparency and enable replication, we further use an open-source capacity expansion model and freely provide the model code and all input data under permissive licenses.

\section{Methods}\label{sec: methods}
We use the open-source power sector model DIETER, which is a linear program that determines least-cost capacity and dispatch decisions for a full year in an hourly resolution \cite{zerrahn_2017,gaete_2021}. DIETER has been used before to study various aspects of renewable integration, electricity storage, and sector coupling \cite{schill_2018,schill_heating_2020,say_2020}. Our application of the DIETER model aims for a level of detail that allows answering the research question without excessively increasing model complexity. DIETER is more detailed in its representation of sector coupling and temporal flexibility than larger integrated assessment models, while being abstract enough to allow for a tractable analysis of a multi-country and multi-sector energy system in an hourly resolution. The model treats the power sector as a copper plate, which means that electricity grid constraints are not considered. Total system costs (all variable, fixed and investment costs of electricity generation, storage, international exchanges and transport) are minimized according to the objective function in equation \ref{eq:objectivefunction}. The energy balance (equation \ref{eq:energybalance}) has to be satisfied for every country $n$ in every hour $h$. It states that the total electric load, electricity storage inflows and electricity requirements for hydrogen production have to be covered by the sum of all domestic fossil and renewable electricity generation, electricity storage outflows and net electricity imports ($F_{l,h}$ denotes exports).

\begin{multline}\label{eq:objectivefunction}
    Z = \sum_n (\sum_h (VARCOST^{fossil}_{n,h} + VARCOST^{RES}_{n,h}\\ 
    + VARCOST^{STO}_{n,h} + VARCOST^{H2_{TRANS}}
        )\\
    + INVCOST^{fossil}_{n} + INVCOST^{RES}_{n} + INVCOST^{STO}_{n} + INVCOST^{H2}_{n}\\ 
    + INVCOST^{H2_{STO}}_{n} + INVCOST^{H2_{TRANS}}_n
    )\\    
    + INVCOST^{NTC}_l + \sum_h VARCOST^{F}_{l,h}
\end{multline}

\begin{multline}\label{eq:energybalance}
    demand_{n,h} + \sum_{sto} STO^{in}_{n,sto,h} + \sum_{h2} E^{H2}_{n,h2,h}
    =\\
    \sum_{tech} G^{fossil}_{n,tech,h} + \sum_{tech} G^{RES}_{n,tech,h} + \sum_{sto} STO^{out}_{n,sto,h}\\ 
    - \sum_l \eta^{ntc}_l \cdot F_{l,h}
\end{multline}

A previous version of the model introduced a detailed representation of hydrogen supply chains for filling stations, including three centralized options (via gaseous hydrogen, liquid hydrogen, or liquid organic hydrogen carriers (LOHC)) and small-scale on-site electrolysis \cite{stoeckl_2021}. In the present study, we use a simplified version of this hydrogen model by excluding liquefied and LOHC hydrogen supply chains and not focusing on filling stations with their particular demand structure, but unspecified industrial hydrogen demand. We focus on on-site and off-site generation of gaseous hydrogen in combination with different types and locations of hydrogen storage (see Figure \ref{fig:h2model}). Key input parameters for the model are capital costs (CAPEX) and operational costs (OPEX) of generation and storage technologies, their efficiencies, and time-series data of availability profiles for non-dispatchable renewable energy sources and load profiles. The model code and input data are available open source.\footnote{Code and data can be found here: \url{https://gitlab.com/diw-evu/projects/modezeen_multi-model}. The documentation of the model can be found here: \url{https://diw-evu.gitlab.io/dieter_public/dieterpy/}}\\

We model Germany and its interconnection with neighboring countries, including Italy. In order to reduce computational complexity and increase tractability, we only allow for endogenous capacity expansion in Germany, while holding the capacity portfolio in other countries constant. That means, that additional electricity demand induced by green hydrogen production in Germany can be satisfied by electricity imports, but cannot induce additional capacity expansion abroad. Over the full year, the model is required to satisfy 80 percent of the non-hydrogen electric load in Germany with domestic renewable electricity generation. This is in line with the German government targets for the renewable energy share in 2030 \cite{BMWK_2022}.In other countries, we do not consider renewable energy targets. The electricity needed for producing hydrogen has to be covered by 100 percent additional renewable energy in a yearly balance. Equation \ref{eq:RES-constraint} ensures that the respective renewable energy target share ($\phi^{RES} = 0.8$) of the system load (except for hydrogen production) is met in Germany. That means that this share of the system load can only be met by energy from biomass, renewable energy sources such as wind, solar and run-off-river hydro power as well as hydro power reservoirs. In energy system models with a binding renewable energy constraint, it can happen that the model transforms renewable curtailment into storage losses by continuously charging and discharging energy storage options \cite{kittel_2022}. In order to avoid this model artifact of unintended storage cycling, we require all electricity losses related to energy storage to be fully covered by renewable electricity in a yearly balance. The final summand of equation \ref{eq:RES-constraint} ensures that electricity for hydrogen production is fully sourced from renewable energy sources.\\

\begin{multline}\label{eq:RES-constraint}
    \sum_h (G^{fossil}_{n,'bio',h} + G^{RES}_{n,tech,h} + STO^{out}_{n,'rsvr',h})
    =\\
    \phi^{RES}_n \cdot \sum_h demand_{n,h}\\
    + \sum_h (\sum_{sto} STO^{in}_{n,sto,h} - STO^{out}_{n,sto,h}
    + \sum_{h2} E^{H2}_{n,h2,h})
\end{multline}

We further assume that hydrogen is only produced and used in Germany, abstracting from evolving hydrogen sectors in neighboring countries. Since we focus on green hydrogen, we force the model to fully cover the electricity demand of electrolyzers and compressors with additional renewable electricity generation in a yearly balance (but not necessarily in every hour). We generally assume that hydrogen is used in the industrial sector, with a flat hourly profile throughout the year. Endogenous model decisions in the hydrogen sector are the investments and hourly use of the two electrolysis processes and the hydrogen storage option.

\section{Data and scenarios}\label{sec: data}
The data for the hourly load profile\footnote{The assumed weather year for the electric load is 1984, following the representative climate years (1982, 1984 and 2007) identified in the TYNDP 2020 \cite{entsoe_2022}.}, net transfer capacities (NTC) between countries, and generation capacities are derived from the TYNDP 2020 scenario ``Distributed Energy'' \cite{entsoe_2022}. According to this scenario, the overall yearly load in Germany is 648~TWh and the peak load is 97~GW in 2030. In the case of Germany, the generation capacities of the TYNDP 2020 are used as lower capacity bounds for renewable energy sources as well as for hard coal and lignite. Additionally, we employ upper capacity bounds on wind capacity expansion in half of the model runs, which are 30~GW for offshore wind and 100~GW for onshore wind. These are inspired by the current German government's plans for wind capacity as of early 2022.\footnote{By the time of publication of this analysis, the German government's onshore wind capacity target has increased to 115~GW.} In contrast, we assume no upper limit for the expansion of solar PV. Table~\ref{tab:capacities} illustrates the upper and lower capacity and electricity storage bounds for Germany, and the assumed fixed capacities in other countries.\\

% Table generated by Excel2LaTeX
\begin{table}[H]
\small
  \caption{Assumed upper and lower capacity bounds in Germany and fixed capacities in other countries, in GW}
  \resizebox{\columnwidth}{!}{
    \begin{tabular}{lrrrrrrrrrrrr}
    \hline
    & \multicolumn{12}{c}{\textbf{Country}} \\
    & \multicolumn{2}{c}{\textbf{DE}} & \multicolumn{1}{c}{\textbf{AT}} & \multicolumn{1}{c}{\textbf{BE}} & \multicolumn{1}{l}{\textbf{CH}} & \multicolumn{1}{l}{\textbf{CZ}} & \multicolumn{1}{l}{\textbf{DK}} & \multicolumn{1}{l}{\textbf{FR}} & \multicolumn{1}{l}{\textbf{LU}} & \multicolumn{1}{l}{\textbf{IT}} & \multicolumn{1}{l}{\textbf{NL}} & \multicolumn{1}{l}{\textbf{PL}} \\
    \hline\\
    \textbf{Technology} & \multicolumn{1}{c}{\textbf{Lower}} & \multicolumn{1}{c}{\textbf{Upper}} & \multicolumn{10}{c}{\textbf{fixed capacities}} \\
    \hline\\
    Run-of-river hydro & 4.04  & 4.04  & 6.14  & 0.15  & 4.11  & 0.40  & 0.00  & 13.64 & 0.05  & 5.64  & 0.05  & 0.54 \\
    Nuclear & 0.00  & 0.00  & 0.00  & 0.00  & 1.19  & 4.04  & 0.00  & 58.21 & 0.00  & 0.00  & 0.49  & 0.00 \\
    Lignite & 7.68  & - & 0.00  & 0.00  & 0.00  & 3.89  & 0.00  & 0.00  & 0.00  & 0.00  & 0.00  & 6.32 \\
    Hard coal & 6.60  & - & 0.00  & 0.62  & 0.00  & 0.37  & 0.77  & 0.00  & 0.00  & 0.00  & 0.00  & 9.88 \\
    Natural gas (CCGT) & 0.00  & - & 2.82  & 7.61  & 0.00  & 1.35  & 0.00  & 6.55  & 0.00  & 38.67 & 8.65  & 5.00 \\
    Natural gas (OCGT) & 0.00  & - & 0.59  & 1.08  & 0.00  & 0.00  & 0.00  & 0.88  & 0.00  & 5.40  & 0.64  & 0.00 \\
    Oil   & 0.00  & - & 0.17  & 0.00  & 0.00  & 0.01  & 0.00  & 0.00  & 0.00  & 0.00  & 0.00  & 0.00 \\
    Other & 0.00  & - & 0.95  & 1.32  & 0.89  & 1.23  & 0.24  & 1.87  & 0.03  & 5.99  & 3.77  & 6.82 \\
    Bio energy & 6.64  & - & 0.60  & 0.21  & 1.20  & 1.06  & 0.67  & 2.56  & 0.05  & 4.93  & 0.54  & 1.41 \\
    Onshore wind & 95.50 & 100.00 & 10.00 & 5.93  & 1.25  & 3.00  & 5.48  & 44.11 & 0.35  & 19.05 & 8.30  & 11.28 \\
    Offshore wind & 17.34 & 30.00 & 0.00  & 4.30  & 0.00  & 0.00  & 4.78  & 3.00  & 0.00  & 0.60  & 6.72  & 0.90 \\
    Solar PV & 109.88 & - & 15.00 & 13.92 & 11.00 & 10.50 & 4.75  & 42.63 & 0.25  & 49.33 & 15.46 & 12.19 \\
    Reservoirs & 2.94  & 2.94  & 7.83  & 0.00  & 8.15  & 1.17  & 0.00  & 10.09 & 0.00  & 13.07 & 0.00  & 0.36 \\
    (Power out)&&&&&&&&&&&&\\
    Lithium-Ion batteries& 0.00  & -     & 0.53  & 0.90  & 0.39  & 0.50  & 0.44  & 3.10  & 0.06  & 1.56  & 0.75  & 0.25 \\
    (Power in/out)&&&&&&&&&&&&\\
    Pumped hydro storage & 8.39  & 8.39  & 5.70  & 1.40  & 3.99  & 1.16  & 0.00  & 3.50  & 1.31  & 11.90 & 0.00  & 1.50 \\
    (Power in/out)&&&&&&&&&&&&\\
    \hline
    \end{tabular}
    }
  \label{tab:capacities}
\end{table}

Prices for fuels and CO$_{2}$ stem from different data sources and are summarized in Table \ref{tab:fuelprices}. The CO$_{2}$ price assumption is based on an analysis by the Potsdam Institute for Climate Impact Research \cite{pietzcker_2021}. Using the REMIND Model, they investigate the required price level of CO$_{2}$ prices in order to achieve the EU climate targets according to the "Fit for 55" package. These imply an emission reduction of 61 percent compared to 2005 in all ETS (Emission trading system) sectors (which includes the power and industry sectors). Assuming that a CO$_{2}$ price is the sole policy instrument in order to achieve this emission reduction, the authors find that this would require a CO$_{2}$ price of 130 Euro per ton in the ETS sectors. Time series data of the availability of renewable energy sources such as onshore wind, offshore wind and solar PV are not provided by \cite{entsoe_2022}, and thus are based on the Open Power System Data platform \cite{opendata,wiese_2019} and provided by \cite{poestges_2022}. Run-of-river water flows are adapted from the same source. Renewable time series are given for the year 2016, which is the base year in the analysis of \cite{poestges_2022}.\\ 

\begin{table}[H]
  \centering
  \caption{Fuel and CO$_{2}$ Prices}
  \resizebox{0.8\columnwidth}{!}{
    %\small
    \begin{tabular}{l l c l l}
    \hline
    \multicolumn{1}{l}{\textbf{Fuel}} & \textbf{Unit} & \textbf{price year} & \textbf{2030} & \textbf{Source} \\
    \hline
    Oil   & EUR/MWh & 2016  & 29.00  & IEA WEO (2020) \\
    Lignite & EUR/MWh & 2016  & 5.50   & NEP 2019 \\
    Hard coal  & EUR/MWh & 2016  & 8.30   & IEA WEO (2020) \\
    Natural gas  & EUR/MWh & 2016  & 14.70  & IEA WEO (2020) \\
    Uranium & EUR/MWh & 2016  & 3.40   & EWI \\
    Bio energy & EUR/MWh & 2016  & 32.50  & BDI (2018) \\
    CO$_{2}$ (EU-ETS) & EUR/ton & 2016  & 130.00  & \cite{pietzcker_2021}
    \\
    \hline
    \end{tabular}
  }
  \label{tab:fuelprices}
\end{table}

Exogenous input parameters for electrolyzers and hydrogen storage options are described in detail in \cite{stoeckl_2021} and summarized in Table \ref{tab:electrolyzers}. The total hydrogen demand of 28~TWh is derived from Germany's national hydrogen strategy, scaled by the new targets set by the current German government for 2030. Assuming a flat demand profile, this yields an hourly hydrogen demand of 3196 MWh in every hour of the year. Since the National Hydrogen Strategy of 2020 associate an electrolyzer capacity of 5~GW with a hydrogen demand of 14~TWh, we assume the updated target of 10~GW to be associated with twice the initial hydrogen demand. It should be noted that electrolyzer capacity is determined endogenously in our model, such that capacity targets only serve as a guideline for our parametrization.\\

\begin{table}[H]
  \centering
  \caption{Exogenous parameters for electrolyzers and hydrogen storage options}
  \resizebox{\columnwidth}{!}{
    \begin{tabular}{lrrrr}
    \hline
          & Investment costs & Annual maintenance costs & efficiency [0,1] & lifetime \\
    \hline
    PEM electrolysis & 724 EUR/kW & 10.86 EUR/kW & 0.71  & 10 years \\
    Alkaline electrolysis & 550 EUR/kW & 8.25 EUR/kW & 0.66  & 10 years \\
    Hydrogen tank storage & 13.50 EUR/kWh & 0.27 EUR/kW & -     & 20 years \\
    Hydrogen cavern storage & 0.10 EUR/kWh & 0.003 EUR/kW & -     & 30 years \\
    \hline
    \end{tabular}
  }
  \label{tab:electrolyzers}
\end{table}

\begin{figure}[H]
    \centering
    \includegraphics[width=\textwidth]{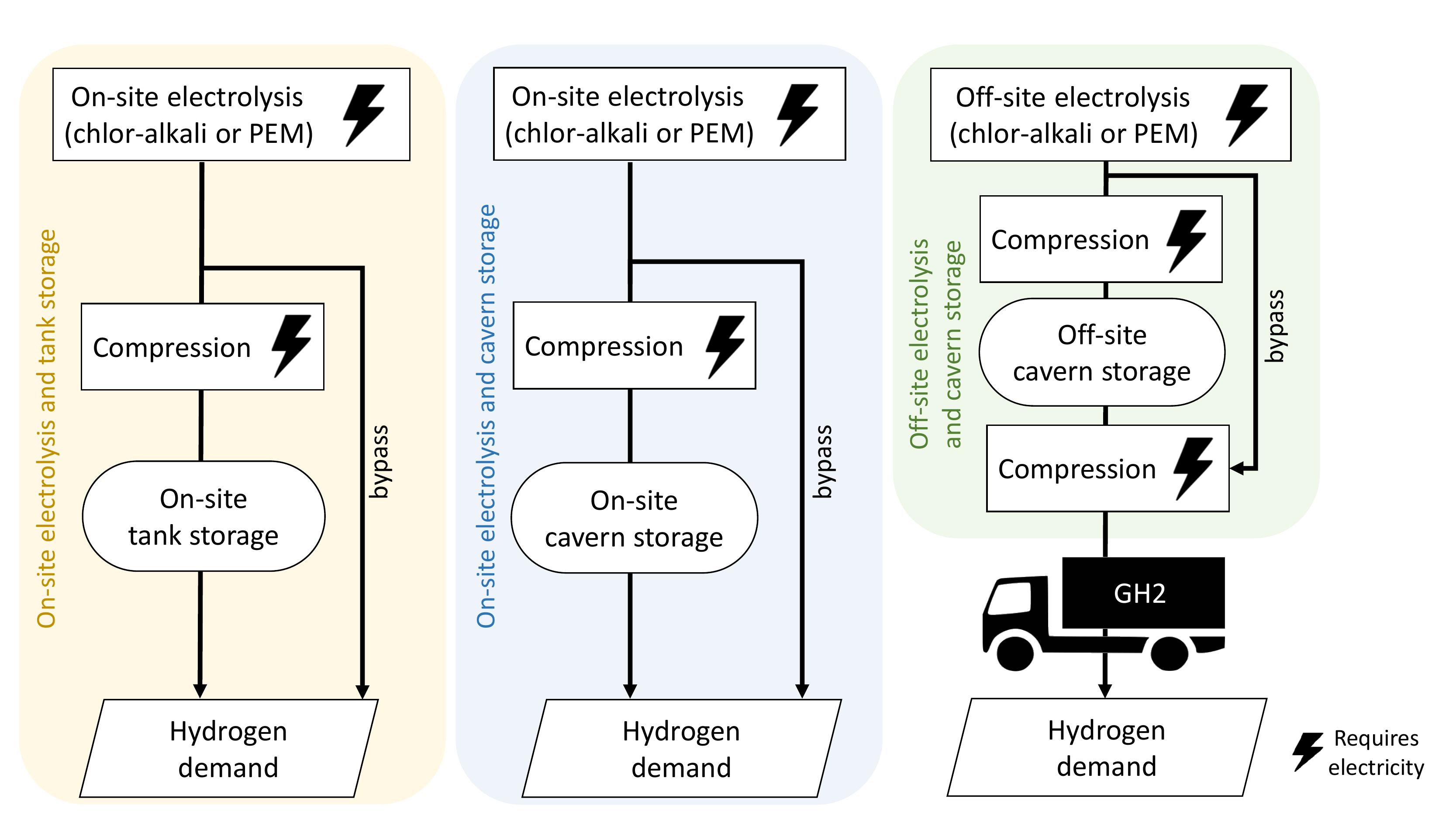}
    \caption{Hydrogen supply infrastructure in the different hydrogen settings}
    \label{fig:h2model}
\end{figure}

We model three stylized hydrogen supply chains, focusing on the effect of different hydrogen storage options (Figure~\ref{fig:h2model}).

Two different electrolysis processes are available for investment: alkaline water electrolysis and proton exchange membrane (PEM) electrolysis. The former has lower investment and operational costs, but a lower efficiency, whereas the latter is more expensive, but also more efficient (\cite{saba_2018}). The analysis focuses on gaseous hydrogen, which means that compression before storage is accounted for. For the sake of simplicity, we do not consider other forms of storing or transporting hydrogen, such as liquefaction or liquid organic hydrogen carriers. In order to clearly separate the effects of the hydrogen storage options on power system outcomes, the three different storage alternatives are mutually exclusive in each model setting. In the first two cases, hydrogen generation and storage is assumed to take place at the sites of hydrogen demand (on-site), whereas the third case assumes that centralized cavern storage options are distant from the locations of hydrogen demand (off-site), such that road transport is required. Tank storage is more expensive than cavern storage; but the latter may incur transport costs. We abstract from a spatial resolution in the hydrogen sector and use an average distance of 250~km between the cavern and the hydrogen demand site in order to estimate transport costs.\\

% Overall number of scenarios / model runs
We consider the combination of two model settings regarding interconnection and wind capacity expansion. First, we differentiate between a setting in which Germany is modeled as an electric island ('DE isolated') and one with interconnection to its neighboring countries ('DE interconnected'). The former helps to identify the general effects of green hydrogen in a power sector with a high share of renewable energy, without any distortions from other regions; the latter, however, is more policy relevant. The differentiation between the two settings also allows drawing conclusions on model distortions related to a limited geographic coverage of energy models. Second, the capacity bounds on wind energy expansion introduced in Table~\ref{tab:capacities} are included in one setting ('Constrained wind') while wind energy expansion is not constrained in another ('Unconstrained wind'). Again, the latter reveals more general effects in a long-run equilibrium perspective; yet the former is more policy-relevant, as the potential for wind power expansion in Germany until the year 2030 is in fact limited, amongst other reasons due to long lead times of admission processes and project development. In addition, we carry out individual model runs for each of the three different hydrogen supply chains introduced above. This allows separating the effects of limited (i.e.,~more expensive) or large-scale (less expensive) hydrogen storage. We accordingly model four reference scenarios without hydrogen, and three hydrogen scenarios to be compared against the respective references. In summary, this results in sixteen distinct model runs.

\section{Results}\label{sec: results}

\subsection{Generation capacity investments and yearly dispatch}

\begin{figure}[H]
    \centering
    \includegraphics[width=1.2\linewidth]{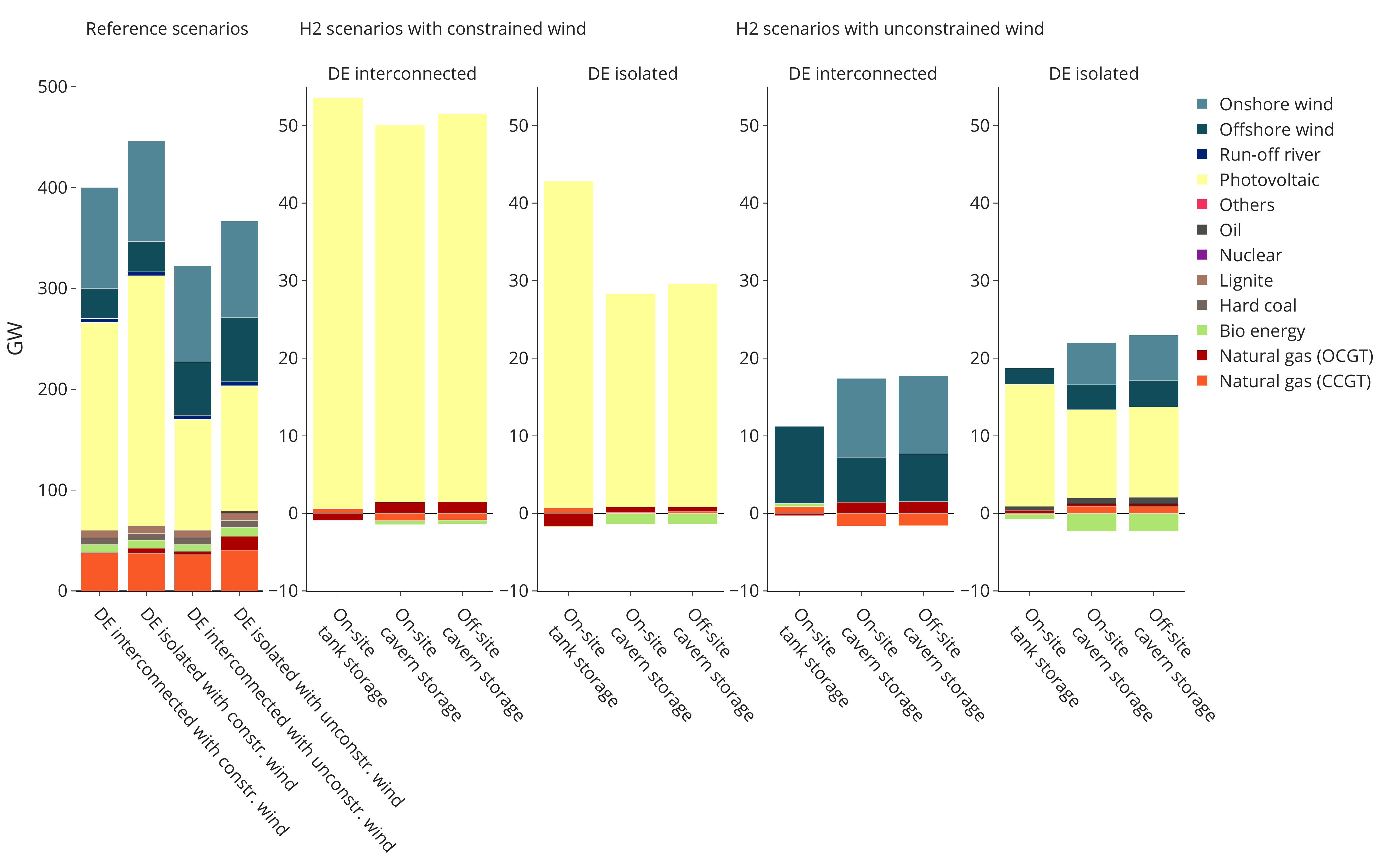}
    \caption{Generation capacity investments in 2030: absolute capacities in reference scenarios (left panel) and changes induced by green hydrogen}
    \label{fig:changes_investments}
\end{figure}

Figure \ref{fig:changes_investments} shows the installed generation capacity in the four reference scenarios (left panel) and the respective changes in the scenarios with green hydrogen production. In the reference scenarios with unconstrained wind expansion, some onshore wind capacity is replaced by cheaper offshore wind capacity, since full-load hours of offshore wind are higher. For the same reason, PV installations also decrease when more (offshore) wind capacity can be built. As a result, offshore wind capacity is about 20 to 30 GW higher than in the scenarios with constrained wind capacities. It can also be seen that international balancing induces a reduction in overall capacity needs. The green hydrogen demand induces additional capacity investments in all model runs. In case wind power cannot be expanded beyond the assumed capacity bound of 100~GW, the additionally required capacity is mainly covered by solar PV, such that between 254~GW and 258~GW of total installed PV capacity are required in the interconnected case, and between 275~GW and 290~GW in the isolated case. Furthermore, the hydrogen supply chain with tank storage requires higher additional capacity investments in solar PV than with cavern storage, because large-scale cavern storage allows for better balancing of seasonal fluctuations in solar energy generation and hydrogen demand. This effect is more pronounced if Germany is modeled as an electric island, since the added flexibility of the caverns becomes even more relevant in the absence of inter-country balancing. However, the opposite is the case if the expansion of wind power is not constrained. Then, green hydrogen triggers additional investments in wind power plants exclusively in 'DE interconnected', accompanied by capacity expansion of solar PV in the case 'DE isolated'. In general, more additional capacity is required in the 'DE interconnected' case, since the total installed capacity in the reference scenario is lower compared to the isolated case and there are more renewable energy surpluses available.\footnote{In the case with unconstrained wind power, this may look differently on first glance (two right-hand side panels in Figure~\ref{fig:changes_investments}); but here, much more wind capacity is deployed in 'DE interconnected' than in 'DE isolated', which has a higher capacity factor than PV.}\\

As for investments into electricity storage, we find only minor effects of green hydrogen (see Figure~\ref{fig:n_sto} in the Appendix). Adding green hydrogen supply chains to the power sector hardly changes the optimal capacities of lithium-ion stationary storage in the parameterization used here. Investments in hydrogen storage capacity (in terms of energy), in contrast, are around two to three orders of magnitude higher than investments in electricity storage.

\begin{figure}[H]
    \centering
    \includegraphics[width=1.2\linewidth]{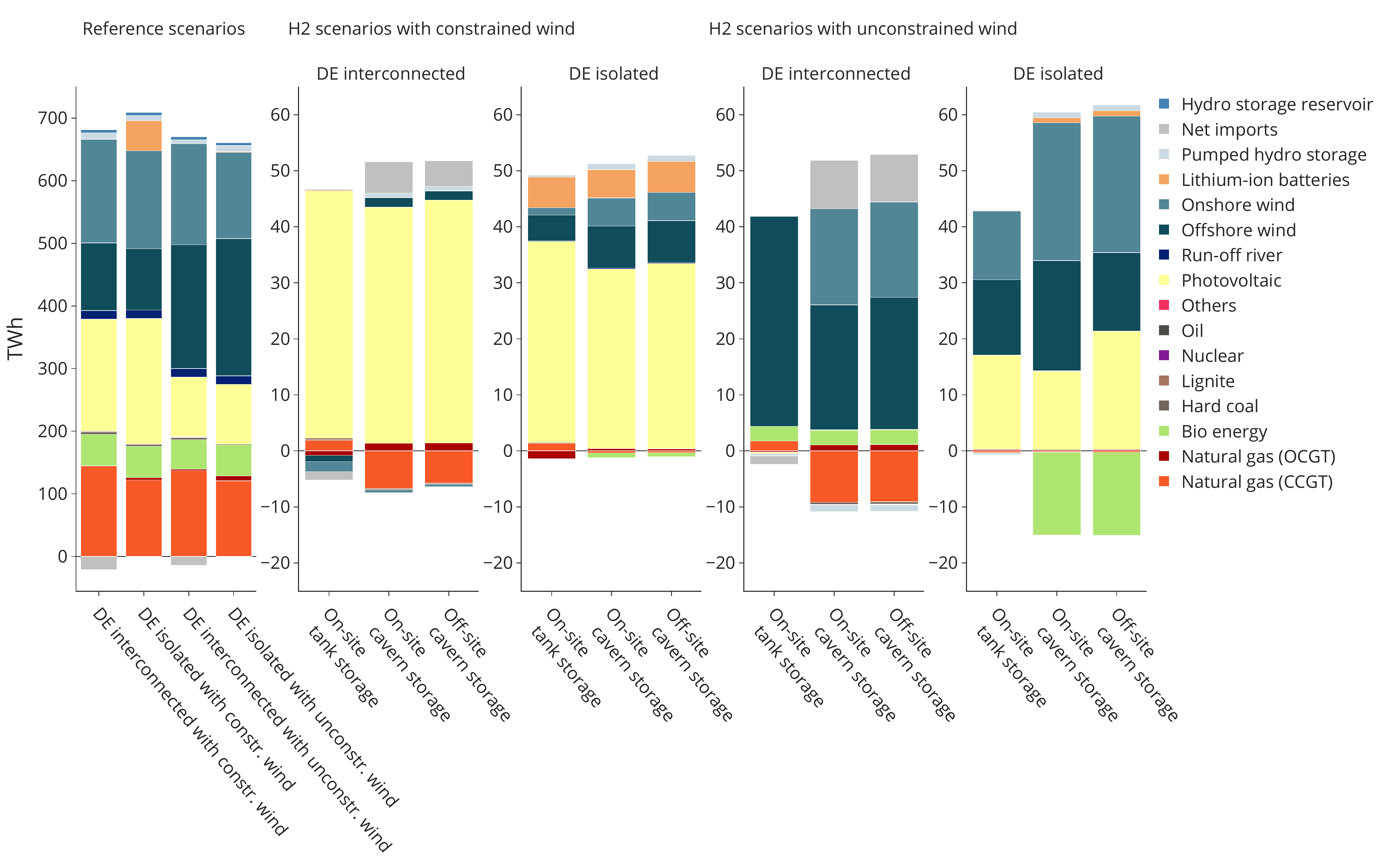}
    \caption{Yearly electricity generation in 2030: absolute values in reference scenarios (left panel) and changes induced by green hydrogen}
    \label{fig:changes_dispatch}
\end{figure}

The total electricity generation over the year is depicted in Figure~\ref{fig:changes_dispatch}, with the absolute values of the reference scenarios in the left panel and changes in yearly dispatch on the right-hand side panels. When wind capacity expansion is constrained, additional electricity demand is mainly met by solar PV, in line with the capacity investments shown above. Yet, there is also an increase in wind power generation, as some wind power is curtailed in the reference. In the interconnected settings, cavern storage (in contrast to tank storage) induces a reduction in net imports to other countries, because renewable surpluses can be captured with the help of these large-scale storage options instead of being exported. Other flexibility options such as pumped hydro storage are also partly substituted by the added flexibility in the hydrogen sector. Furthermore, the additional flexibility of cavern storage also leads to a reduction in fossil fuel generation, mainly from natural gas and a bit from coal. In the isolated settings, the operation of short-duration electricity storage is required to compensate renewable surpluses, even in the presence of large-scale cavern storage. When wind energy expansion is unconstrained, these electricity storage needs become less relevant, since offshore wind provides much more electricity, particularly if only on-site tank storage of hydrogen is available. Offshore wind has a much flatter generation profile than onshore wind, which reduces the need for seasonal storage and outweighs its higher levelized costs of electricity in this case. If Germany is modeled in isolation, we find some additional solar PV generation even in the case of unconstrained wind power expansion, as this leads to fewer renewable energy surpluses in this setting where surpluses cannot be balanced with other countries.\\

\begin{figure}[H]
    \centering
    \includegraphics[width=\textwidth]{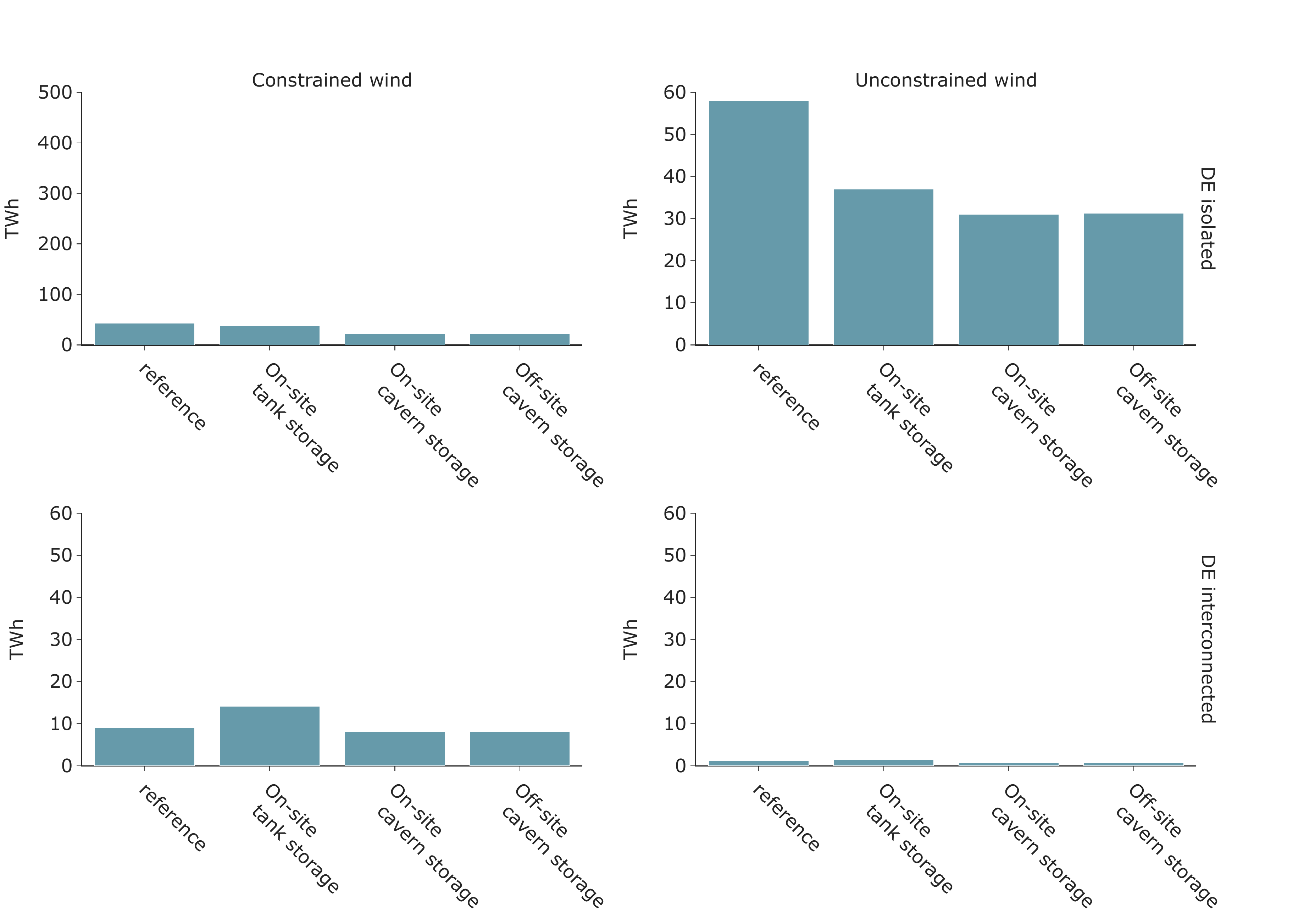}
    \caption{Yearly renewable curtailment}
    \label{fig:curtailment}
\end{figure}

In nearly all cases, the curtailment of renewable energy decreases after green hydrogen is introduced (Figure~\ref{fig:curtailment}). Note that this happens despite the fact that green hydrogen triggers an additional increase in renewable capacity expansion. Yet, this is not true if only tank storage is available and Germany is modeled in interconnection with its neighbors, particularly if wind power is constrained (lower left panel of Figure~\ref{fig:curtailment}): here, renewable curtailment increases. That is because, with a view to renewable curtailment, the temporal flexibility provided by electrolyzers and tank storage is outweighed by the increase in system-wide flexibility needs which stem from additional variable renewable capacity expansion that is required to cover the electrolyzers' energy demand. Renewable curtailment is further much higher if the German power sector is modeled in isolation (43~TWh and 58~TWh in the reference cases), as compared to an interconnected power system (9~TWh and 1.25~TWh in the reference cases). This is because there is no balancing effect of cross-border transmission on variable renewable and load profiles (see~\cite{schlachtberger_2017,roth_2023}) in the 'DE isolated' setting, i.e. fluctuations cannot be offset by power imports or exports. Thus, renewable generation capacities are higher in this scenario.\\ 

Note that the amount of renewable energy curtailed over the year in this setting is in the order of magnitude, or even higher, as the additional electricity demand of green hydrogen production. The amount of renewable curtailment is lowest if wind power expansion is unconstrained and Germany is interconnected with its neighbors. Here, there is the highest degree of flexibility in the system and the portfolio is optimized such that variable renewable generators can be efficiently dispatched with hardly any curtailment.\\

\subsection{Electrolysis capacity and hourly use}

\begin{figure}[H]
    \centering
    \includegraphics[width=\linewidth]{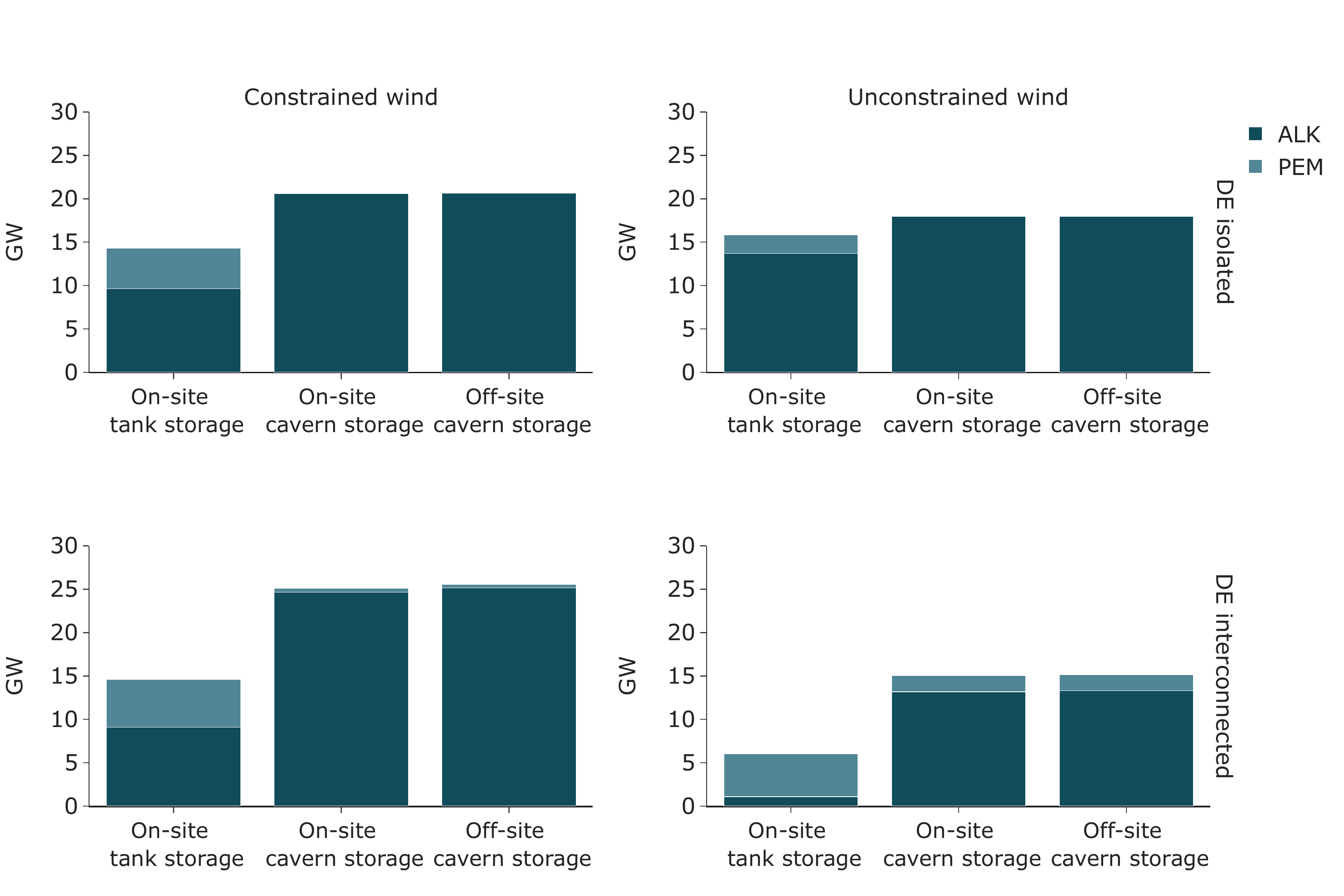}
    \caption{Installed electrolyzer capacity}
    \label{fig:electrolyzer_cap}
\end{figure}

We generally find much higher investments in alkaline (ALK) electrolysis than in the more efficient PEM technology (Figure~\ref{fig:electrolyzer_cap}). This is driven by the fact that renewable electricity is relatively cheap in our parameterization, such that the costs of electricity matter less than the CAPEX of electrolyzers. This finding resembles the one in \cite{stoeckl_2021}. The optimal electrolysis capacity strongly depends on the availability of low-cost cavern storage. Electrolyzer capacity is generally much higher if caverns can be built, as these allow storing hydrogen cheaply in large quantities and thus to temporally decouple electrolyzer operations from hydrogen demand. More specifically, caverns enable electrolysis to draw more electricity from the grid in hours of high renewable availability, i.e.,~when electricity market prices are low. To make most use of these hours, the optimal electrolysis capacity thus increases compared to the case where only tank storage is available and where a more balanced electrolyzer operation profile is optimal. Note that renewable surplus generation occurs in a limited numbers of hours throughout the year; but if they do, they often come with a very high power rating. This means the right-hand side of the residual load duration curve becomes substantially negative with increasing shares of variable renewable energies (see~\cite{schill_2020}). Accordingly, electrolyzers have lower full-load hours in the cases with cavern storage compared to the setting with tank storage (see Figure~\ref{fig:flh} in the Appendix).\\

In most cases, optimal electrolyzer capacities are higher if wind power expansion is constrained as compared to the case where wind power can be deployed without limits. This is again because a wind-constrained system has a higher PV capacity. This causes large temporal renewable surplus generation around midday especially in summer, which is absorbed with a higher electrolyzer capacity in the optimum. We further find the highest optimal electrolysis capacity if Germany is modeled in interconnection with its neighbors, and wind power is constrained - despite the fact that overall yearly renewable curtailment is much higher in the 'DE isolated' case.\\

\begin{figure}[H]
    \centering
    \subfloat{\includegraphics[valign=t,width=0.5\linewidth]{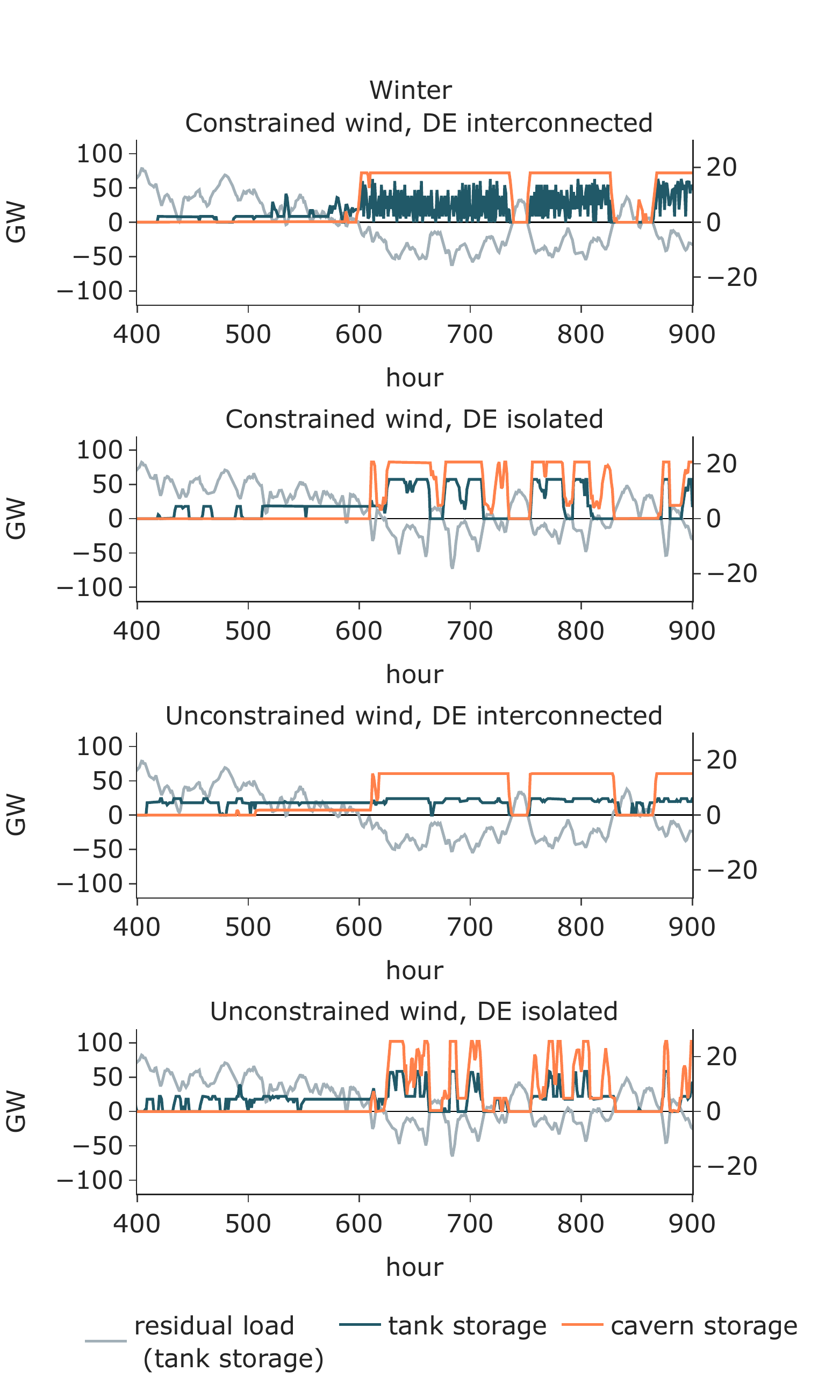}}
    \subfloat{\includegraphics[valign=t,trim=0 50 0 0, clip, width=0.5\linewidth]{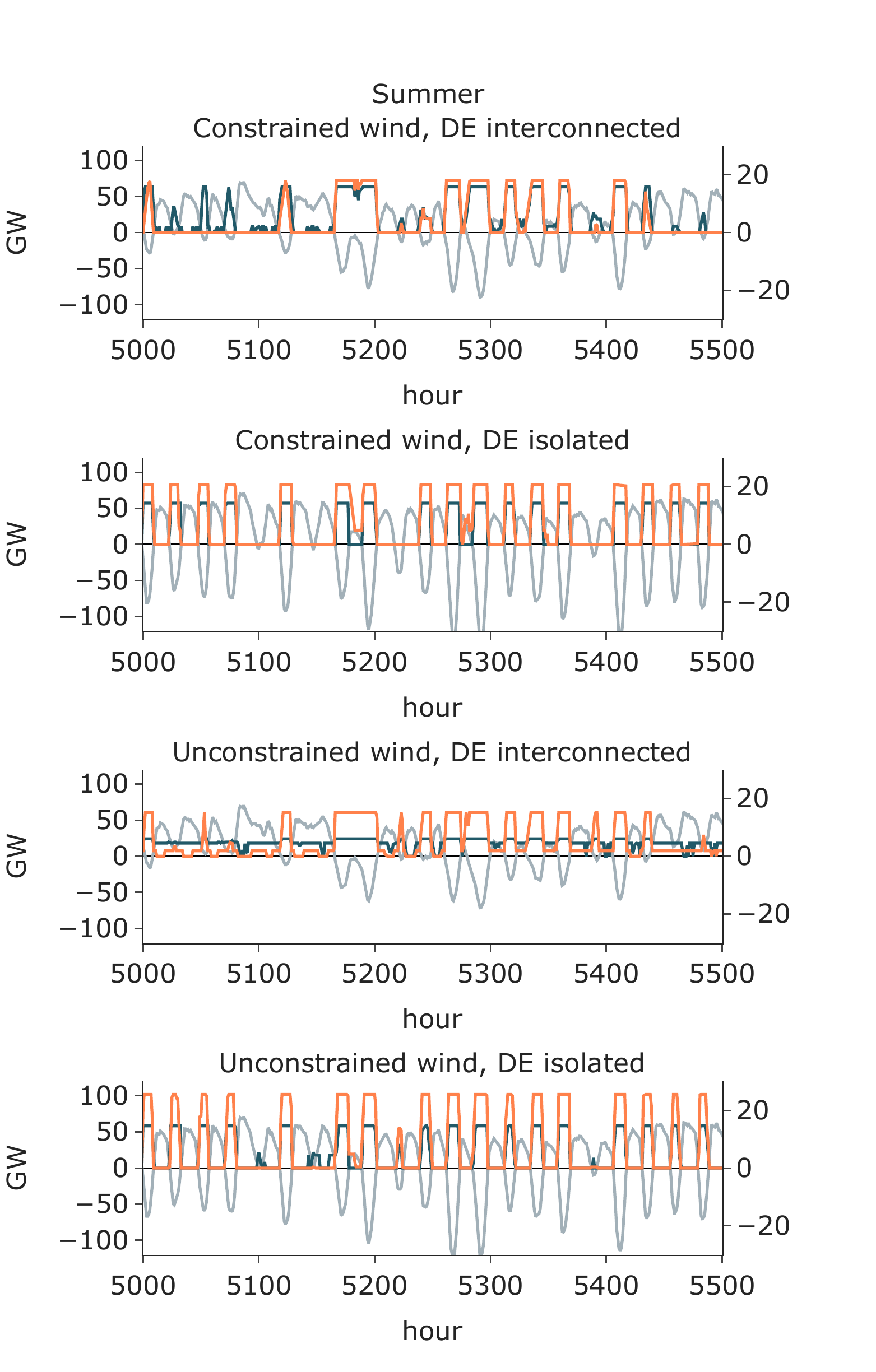}}
    \caption{Electricity consumption of electrolysis versus residual load in 500 winter hours (left panels) and 500 summer hours (right panels)}
    \label{fig:resload_h2e}
\end{figure}

Figure~\ref{fig:resload_h2e} shows the hourly electricity use by electrolyzers in the 500 exemplary winter and summer hours of the modeled year. Cavern storage, in combination with higher electrolysis capacity, enables a higher use of electrolyzers in periods of renewable surplus generation, i.e., low-price periods, as compared to tank storage. Likewise, hydrogen supply chains with low-cost cavern storage allow minimizing the electricity consumption by electrolyzers in periods of positive residual load, i.e., periods with higher prices. In contrast, electrolyzers that are combined with more expensive tank storage are forced to use more electricity in such periods. In the interconnected setting, a constraint on wind power expansion further leads to a more fluctuating use of tank storage as compare to the case where wind power is unconstrained. This is because solar PV capacity is higher in the wind-constrained case, giving rise to more diurnal variability. Figure~\ref{fig:resload_h2e} also illustrates that renewable surpluses are larger in the isolated setting than in the interconnected one. This is especially visible in the exemplary summer hours, as a large portion of overall renewable surplus generation happens in summer.\\

\subsection{Costs and electricity market prices}

\begin{figure}[H]
    \includegraphics[width=\linewidth]{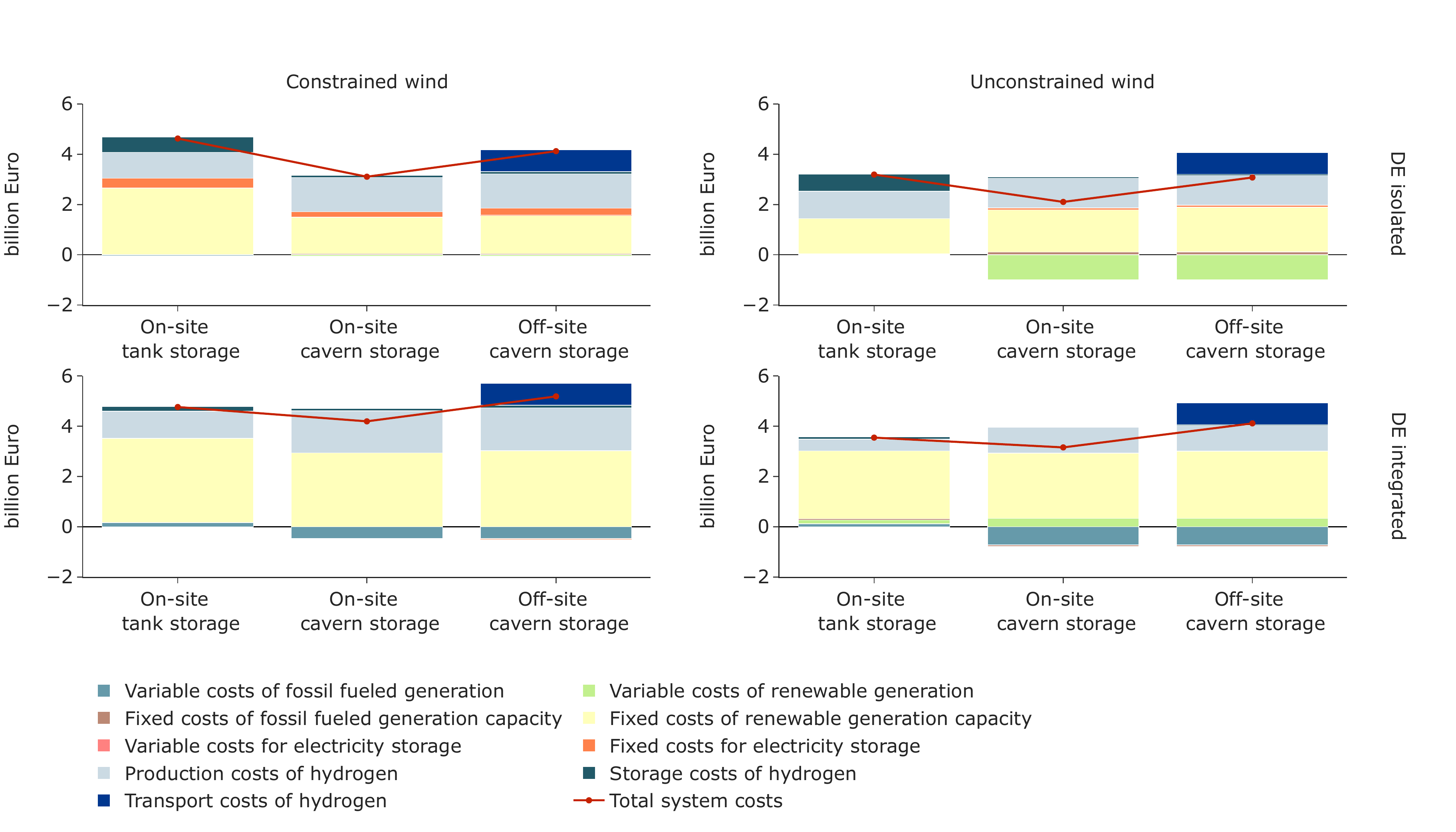}
    \caption{Changes in power sector costs induced by green hydrogen compared to reference scenarios, differentiated by components}
    \label{fig:cost_diff}
\end{figure}

We evaluate the power sector cost increase related to green hydrogen by comparing the results of the model runs with different hydrogen supply chains to the reference without hydrogen (Figure~\ref{fig:cost_diff}). In all modeled cases, overall power sector costs are lowest if investments in cavern storage are possible at the site of hydrogen demand (and production). As shown above, this option provides low-cost temporal flexibility to the system, which allows cost-saving capacity and operation decisions. If, in contrast, only higher-cost tank storage for hydrogen can be built, the power sector costs effects of green hydrogen are higher, as the hydrogen supply chain offers less temporal flexibility. The same is true if the location of cavern storage does not coincide with the location of hydrogen demand. In this case, additional transport costs may outweigh the flexibility benefits of cavern storage. The increase in power sector costs caused by green hydrogen is generally lower if there is no constraint to wind power capacity expansion (right-hand side panels of Figure~\ref{fig:cost_diff}). Green hydrogen further leads to lower increases of power sector costs if Germany is modeled in isolation, as compared to a setting where Germany is interconnected with its neighboring countries, which is caused by larger, and more under-utilized, generation capacities in the more flexibility-constrained 'DE isolated' setting.\\ 

Cost effects are driven to a large extent by additional fixed costs of renewable generation capacity. This is particularly pronounced in the 'DE interconnected' setting, as less renewable surplus generation is available here, and more renewable generation capacity has to be built to cover the additional electricity demand by electrolyzers and compressors. Production costs of hydrogen, which includes all capital and operational costs of electrolysis and compression before storage, are another major cost component. These are generally higher for supply chains with cavern storage, since they are characterized by higher installed electrolyzer capacities compared to supply chains with tank storage (see Figure \ref{fig:electrolyzer_cap}). When Germany is interconnected with its neighbors, hydrogen supply chains with cavern storage further allow reducing the variable costs of fossil fueled generators.\\

\begin{figure}[H]
    \centering
    \includegraphics[width=\textwidth]{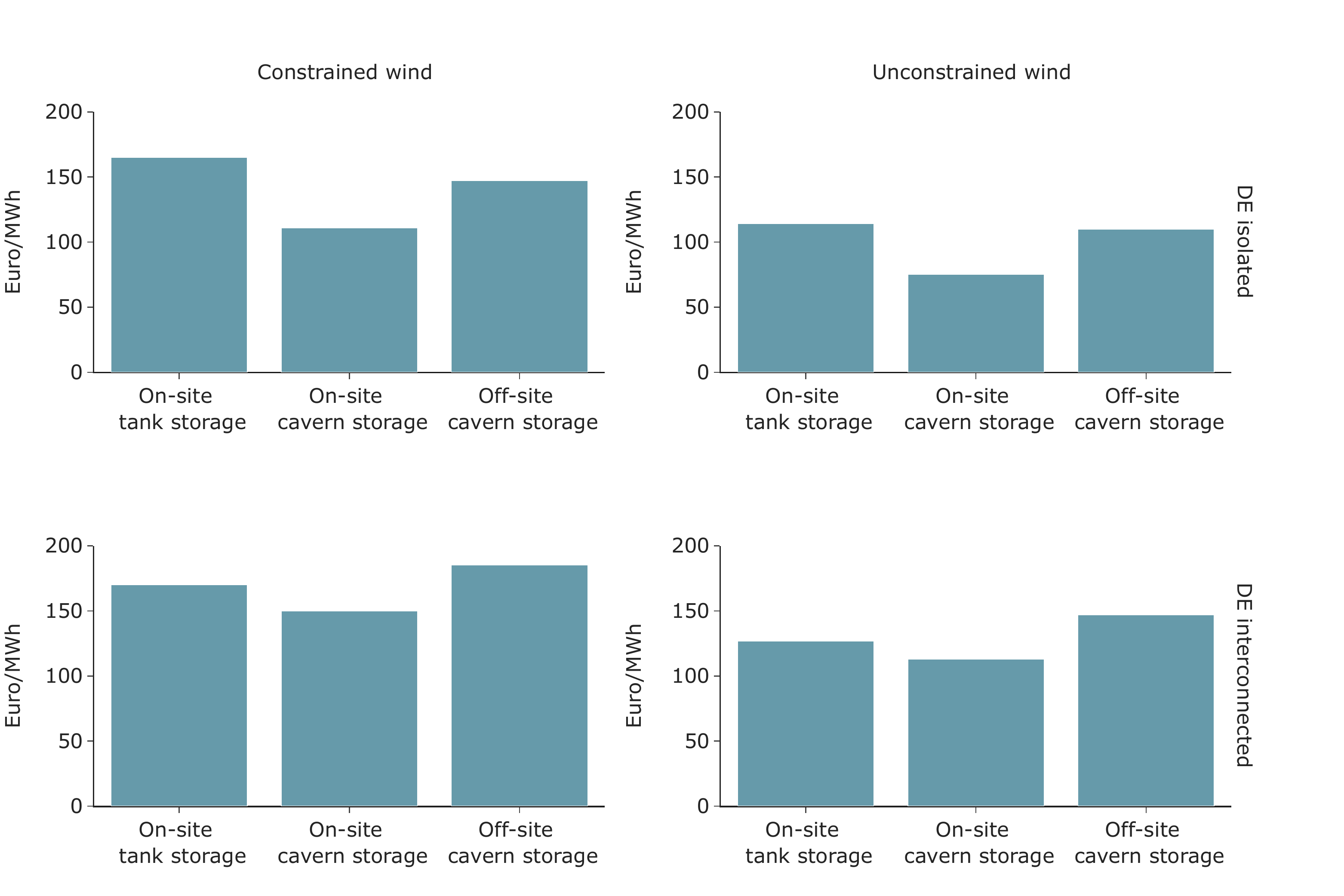}
    \caption{Additional system costs of hydrogen (ASCH)}
    \label{fig:asch}
\end{figure}

If we relate the power sector costs effects to the amount of green hydrogen generated, we get the additional system costs of hydrogen (ASCH \footnote{ASCH = Additional power system costs / Total hydrogen demand}, Figure~\ref{fig:asch}). These have been defined by \cite{stoeckl_2021}, adopting a perspective of additionality of green hydrogen. As a consequence of the cost effects discussed above, we find the lowest system costs of hydrogen in the case where Germany is modeled in isolation and without constraints to wind power expansions, and particularly if cavern storage can be built at the site of hydrogen demand. Here, ASCH are particularly low at around 7.50 Euro-cent/kWh (or 2.50 Euro/kg $H_2$ \footnote{1 kg of $H_2$ contains 33.33 kWh of energy (lower heating value).}), as electrolyzers can make substantial use of otherwise curtailed renewable surplus generation. In contrast, we find the highest ASCH at 18.53 Euro-cent/kWh (6.17 Euro/kg $H_2$) if Germany is modeled in the interconnection with its neighbors, and if the location of hydrogen demand does not coincide with the sites where cavern storage can be built. Here, much less renewable surplus energy is available for electrolyzers, and transport costs outweigh the flexibility benefits of cavern storage.\\

\begin{figure}[H]
    \centering
    \includegraphics[width=\textwidth]{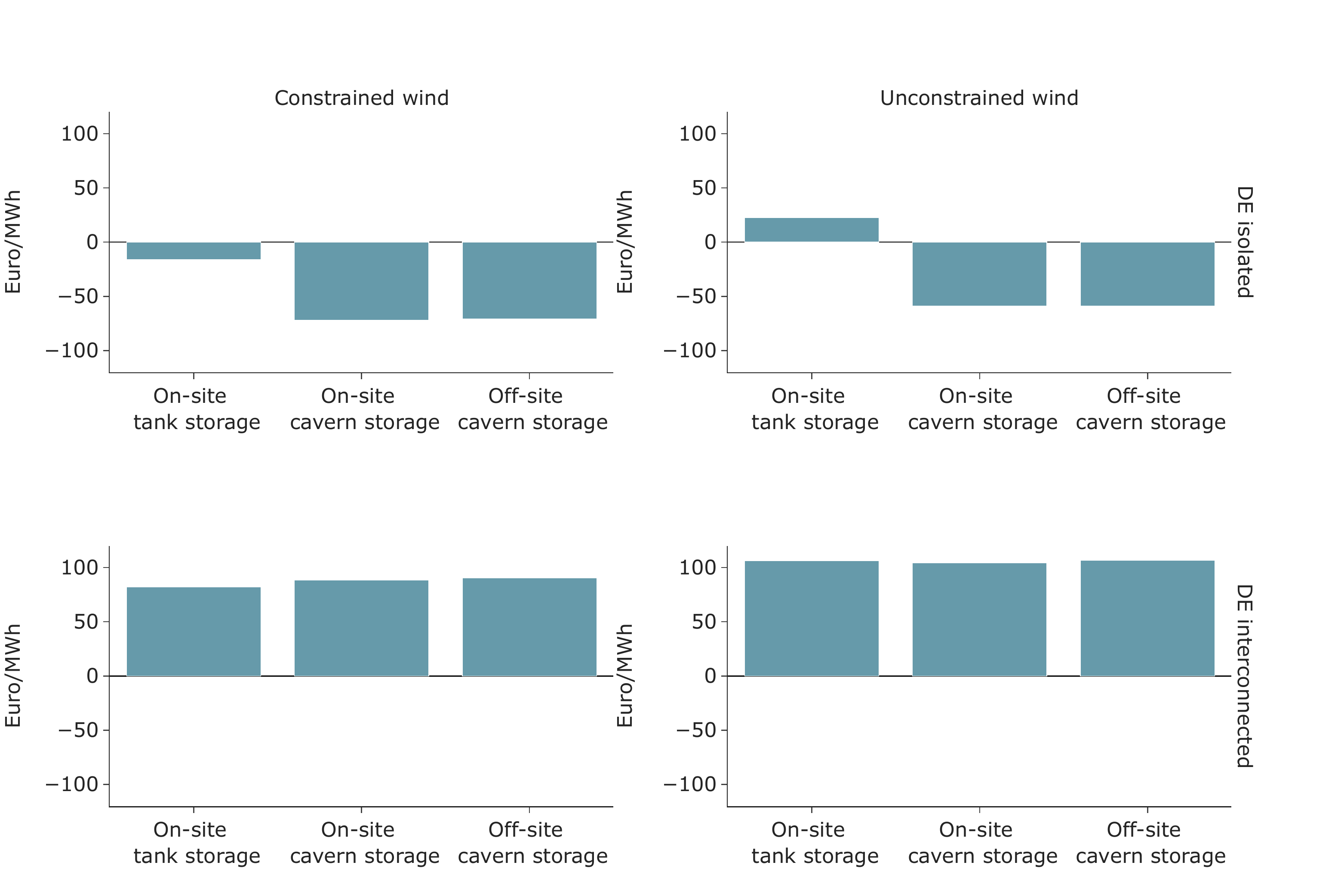}
    \caption{Average provision costs of hydrogen (APCH)}
    \label{fig:apch}
\end{figure}

Next, we look at a complementary metric, the average provision costs of hydrogen (APCH \footnote{APCH = (Annualized costs of hydrogen infrastructure + variable hydrogen costs + yearly electricity bill of hydrogen production) / Total hydrogen supply}, Figure~\ref{fig:apch}). These have also been defined in \cite{stoeckl_2021}. In contrast to the ASCH metrics, the APCH concept adopts a hydrogen producer perspective. It is calculated as the sum of fixed and variable costs related to electrolysis, hydrogen storage and transport, plus the energy bill faced by operators of electrolyzers and compressors. This metric is conceptually closely related to Levelized Costs of Hydrogen (LCOH), which are defined as the net present value of the total investment and operating costs of an electrolyzer divided by the total hydrogen production over its lifetime. Yet, APCH allow for a better comparison to ASCH. Average provision costs of hydrogen are generally lower than additional system costs of hydrogen. Strikingly, they are even negative in most cases where Germany is modeled in isolation, because electrolyzers largely use renewable surplus energy which comes with negative prices. These negative prices are driven by the renewable energy constraint in the model  (\cite{brown_2021, lopez_prol_2021}). APCH are higher, and positive, if Germany is interconnected with its neighbors, but still lower than ASCH. This indicates that hydrogen producers not necessarily internalize the full costs that they cause in the power sector.

\begin{figure}[H]
    \centering
    \includegraphics[width=\linewidth]{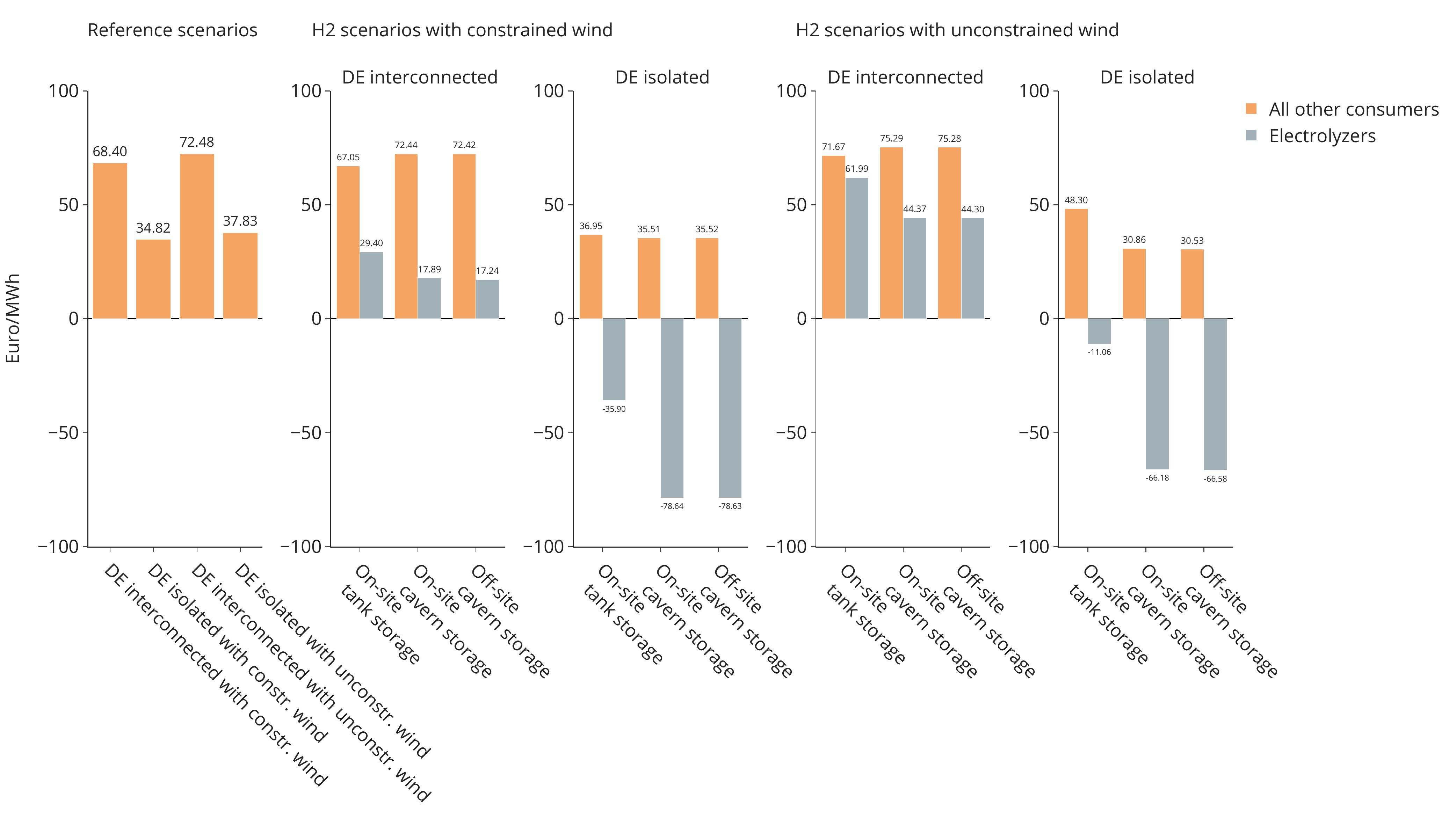}
    \caption{Load-weighted average electricity market prices paid by electrolyzers and all other electricity consumers in Germany (absolute values)}
    \label{fig:pricechanges_h2}
\end{figure}

An analysis of average electricity prices corroborates this finding. Figure~\ref{fig:pricechanges_h2} shows the wholesale electricity prices (average over the full year) paid by electrolyzers and by all other electricity consumers in the reference scenario and the different hydrogen scenarios.\footnote{Note that, because of the binding yearly renewable share constraint in the model, the prices shown here do not cover the full power sector costs. Renewable generators require an additional payment according to the dual variable of the renewable energy constraint in order to achieve zero profits in the long run. This can be interpreted as the payment of an energy-based renewable energy support scheme via quotas or market premiums (see~\cite{brown_2021}). How these payments are financed, e.g.,~via energy surcharges, or via the national budget, is another question and may alter the distributive effects between different types of energy consumers.} In the reference cases without green hydrogen, average prices are much lower if Germany is modeled in isolation as compared to an interconnected setting. This is because achieving the renewable energy constraint in Germany requires a much larger renewable power plant fleet if Germany is modeled as an island. Accordingly, there are higher renewable surpluses and more hours with very low prices.\\

We further find that the average prices of electricity used for electrolysis are generally lower than average prices in corresponding settings without green hydrogen. Accordingly, producers of green hydrogen can use the temporal flexibility potential of their assets to exploit periods with low prices. They can do so to a higher degree if low-cost caverns can be built, visible in lower average prices than in the cases of tank storage. Average electricity prices of electrolyzers are even lower if Germany is modeled in isolation. Here, average prices even become negative, which means that flexible electrolyzers would be paid for the electricity they consume. Note that this finding depends on a strict and binding renewable energy target, which also requires that the yearly electricity demand of electrolyzers has to be fully covered by renewable energies, as well as on the implicit assumption that there are no other flexibility options available except for electricity storage. In such a setting, the dual variable of the energy balance, which we interpret as a wholesale price, can become negative. We can interpret negative prices as an energy-based support payment that a renewable generator needs per MWh in order to not incur losses from production (i.e. to achieve zero profits in equilibrium). Alternatively, we can interpret a negative price as the cost saving that an increase in demand of one MWh in the particular hour would facilitate, considering that this would make it easier to achieve the yearly renewable energy constraint. This interpretation has been described, for example, by \cite{brown_2021} and \cite{lopez_prol_2021}. If Germany is modelled in interconnection with its neighbors, which adds a lot of temporal flexibility to the system, negative average prices for hydrogen electricity vanish.

\section{Discussion}\label{sec: discussion}

In the following, we briefly discuss some limitations and implications of our research design with respect to our assumptions of the power sector and the hydrogen sector, and how these may qualitatively affect our numerical results.\\ 

In the power sector, we do not consider other sector coupling options such as electric mobility or power-to-heat, which may compete with green hydrogen supply chains for renewable energy surpluses. Our findings have shown that the flexibility provided by the European interconnection lowers the value of the flexibility of hydrogen storage options. More flexible electricity demand could thus decrease the marginal flexibility benefits of hydrogen storage, since less renewable surplus energy would be available for green hydrogen production. So far, studies have shown that the combination of different sector coupling options increases power system benefits and can be even more cost efficient than inter-country transmission \cite{brown_2018}. Gils et al.~specifically find that electrified heating can complement flexible green hydrogen production \cite{gils_2021}. In their study, electric heating makes use of power generation peaks in winter, while electrolysis runs during the rest of the year. However, \cite{brown_2018} also highlight that the electrification of flexible transport and heating, in conjuncture with long-term thermal energy storage, crowds out the flexibility provided by electricity storage. It is unclear whether this effect also arises among different sector coupling options and green hydrogen, if renewable energy surpluses are scarce, so further research is required. Moreover, we do not analyze the spatial aspects of domestic hydrogen provision in this study. Studies such as \cite{vomscheidt_2022} suggest that green hydrogen production could enhance bottlenecks in the German electricity grid if there are no adequate spatial price signals. The transmission grid may be particularly congested in settings where electrolyzers are located close to hydrogen demand \cite{lieberwirth_2023}. Transmission bottlenecks may thus impede the flexible use of electrolysis and limit the system benefits of flexible green hydrogen identified in this study.\\    

In the hydrogen sector, we assume that the hydrogen demand in Germany is met solely by domestic green hydrogen provision. Thus, we do not consider potential future imports. Yet, other studies have shown that Germany is expected to import a major share of its hydrogen needs \cite{husarek_2021,gils_2021,brandle_2021}, which is also acknowledged by the German government \cite{BMWI_2020}. If domestic production of hydrogen would be lower due to higher imports, the power sector effects were smaller than identified here. In general, it is questionable whether even the assumed domestic production capacity of 10 GW is achievable by 2030. Odenweller at al.~find that green hydrogen supply will probably remain scarce until 2030 and that unconventionally high growth rates are required to achieve the electrolysis capacity targets within the European Union \cite{odenweller_2022}. It should also be noted that the differences in results between the settings with on-site and off-site cavern storage greatly depend on the assumed transport distance. Transport costs increase with transportation distance and thus the benefits of very remote cavern storage diminish. Furthermore, we only consider road transport of gaseous hydrogen. However, studies have shown that transporting very high volumes of hydrogen would be more cost-efficient via pipelines \cite{reuss_2019}. It has even been proposed that the repurposed natural gas network in Spain could be used for seasonal hydrogen storage \cite{brey_2021}. Considering pipeline transmission would greatly decrease the transport costs and thus lower the system costs of the settings with off-site cavern storage compared to our findings. Furthermore, we only consider wholesale market revenues for electrolyzers, where they benefit from their flexibility by utilizing periods of low electricity prices. However, electrolyzers could also provide ancillary services. This opens up an additional stream of revenue, potentially lowering the costs of green hydrogen production. It has been shown that electrolyzers can profitably participate in the balancing market by providing downward reserves \cite{pavic_2022}. Whether this is economically viable depends not only on techno-economic assumptions on electrolyzers, but also on the demand and supply on future ancillary services markets, as well as respective regulation. In any case, the size of ancillary services markets is likely to remain small relative to wholesale markets. Another aspect of the hydrogen supply chain that we have abstracted from is the reconversion of hydrogen to electricity. It is not clear \textit{per se} whether the inclusion of reconversion would create synergies with green hydrogen provision for industrial demand, or rather compete with it. On the one hand, both demands could make use of the same hydrogen storage capacity. On the other hand, we have shown that electrolyzers tend to use hours with excess renewable energy. When electrolyzers are forced to run in more expensive hours more frequently due to the additional hydrogen demand for reconversion, hydrogen production would become more expensive. The interplay of hydrogen generation for reconversion and industrial demand are thus subject to future research.

\section{Conclusions and policy implications}\label{sec: conclusions}

In this study, we provide a comprehensive picture of the power sector effects of alternative production and storage options for green hydrogen in a mid-term future scenario for Germany. We find that the generation of green hydrogen generally increases the power system cost in all cases modeled here. Accordingly, the additional demand for renewable energy generation related to hydrogen production always outweighs its additional flexibility potentials. Still, we find that cavern storage of hydrogen provides valuable temporal flexibility as it provides seasonal storage in settings with high shares of variable renewable energy sources. This confirms previous findings, e.g.~by \cite{michalski_2017}. Yet, additional transport costs reduce these benefits if caverns are located distantly from hydrogen demand. Especially in the case of less flexible supply chains with tank storage, the additional renewable capacity that is required to cover the electricity demand for green hydrogen may lead to an additional need for flexibility which surpasses the flexibility benefits that the hydrogen sector can provide. Furthermore, flexibility benefits from domestic green hydrogen production are less pronounced within an interconnected power system in which geographical balancing provides a substantial amount of additional flexibility.\footnote{Similar findings hold for the interplay of electricity storage and interconnection \cite{roth_2023}.} Thus, the power system effects of green hydrogen seem to depend greatly on the availability of other flexibility measures, which should also apply to other sector coupling options.\\

We further find that the effects of green hydrogen production on the optimal generation portfolio strongly depend on the model assumptions: if there are no technology-specific constraints on renewable expansion, hydrogen triggers an additional expansion of wind power. If the potential for wind power expansion is limited, green hydrogen can also be supplied with additional solar PV capacity at only moderately higher costs. Our analysis also shows that the additional system costs of hydrogen (ASCH)  are higher than the average provision costs of hydrogen (APCH). This indicates that hydrogen producers may not face the full costs that they cause in the power sector. We further identify a potential unintended distributional effect of green hydrogen, since hydrogen producers may profit from lower electricity prices at the detriment of other electricity consumers.\\

Several policy implications can be derived from our results. First, even a moderate green hydrogen demand of 28~TWh induces a sizeable additional need in renewable energy capacity, which is why policy makers should further push towards a fast capacity expansion of renewable energies. Second, our results show that a higher temporal flexibility of green hydrogen supply chains is desirable from a power system perspective. Thus, policy makers should work towards supporting such flexibility, for example by enabling the expansion of cheap large-scale hydrogen storage options wherever possible. Third, policy makers should enable hydrogen producers to realize their flexibility potentials in their operational decisions, for example by non-distorting regulation and tariff design. At the same time, they should take into account the potential distributional effects of hydrogen production on other types of electricity consumers. While hydrogen producers should be enabled to internalize some of the flexibility benefits they provide to the power sector, detrimental effects on other, and potentially less flexible, consumers should be minimized. Finally, our results further corroborate the well-established finding that large-scale spatial balancing is desirable in power sectors with high shares of variable renewable energy sources (see~\cite{schlachtberger_2017}). We extend this finding to the integration of green hydrogen production. Policy makers should thus continue and increase their efforts to strengthen the common European electricity market and expand cross-border transmission capacities.\\

While our analysis focuses on a mid-term future scenario for Germany, general results should also hold for other countries in the temperate climate zone with comparable load and renewable patterns. Carrying out similar analyses for other countries and world regions, ideally in combination with other options for power sector flexibility and sector coupling, appears to be a promising avenue for future research.

\section*{Acknowledgements}
We thank three anonymous reviewers and the special issue editors for insightful comments. We further thank Marco Breder, Michael Bucksteeg, Hannes Hobbie, Martin Lieberwirth and Felix Meurer  as well as the participants of the online workshop ``Power sector implications of green hydrogen in Germany'' on 7~April~2022, the ``Infrastructure Policy in the 21st Century'' conference at TU~Berlin on 23~September~2022, and the 16th~ENERDAY~conference at TU~Dresden on 30 September 2022 for valuable comments on earlier drafts. We also thank our colleagues Carlos Gaete Morales and Alexander Roth for valuable input and modeling support. We gratefully acknowledge funding by the German Federal Ministry for Economic Affairs and Climate Action (BMWK) via the research project Modezeen, Fkz 03EI1019D, and by the Einstein Foundation via the project Open Hydrogen Modeling, A-2020-612.

\section*{Author contributions}
\textbf{Dana Kirchem}: Methodology; Investigation; Data curation; Software; Visualization; Writing - original draft; Writing - review and editing; \textbf{Wolf-Peter Schill}: Conceptualization; Methodology; Investigation; Writing - original draft; Writing - review and editing; Funding acquisition.

\appendix
\setcounter{figure}{0}

\section{Complementary model results}

\subsection{electricity and hydrogen storage capacity}\label{sec: appendix - storage}

\begin{figure}[H]
    \centering
    \includegraphics[width=\linewidth]{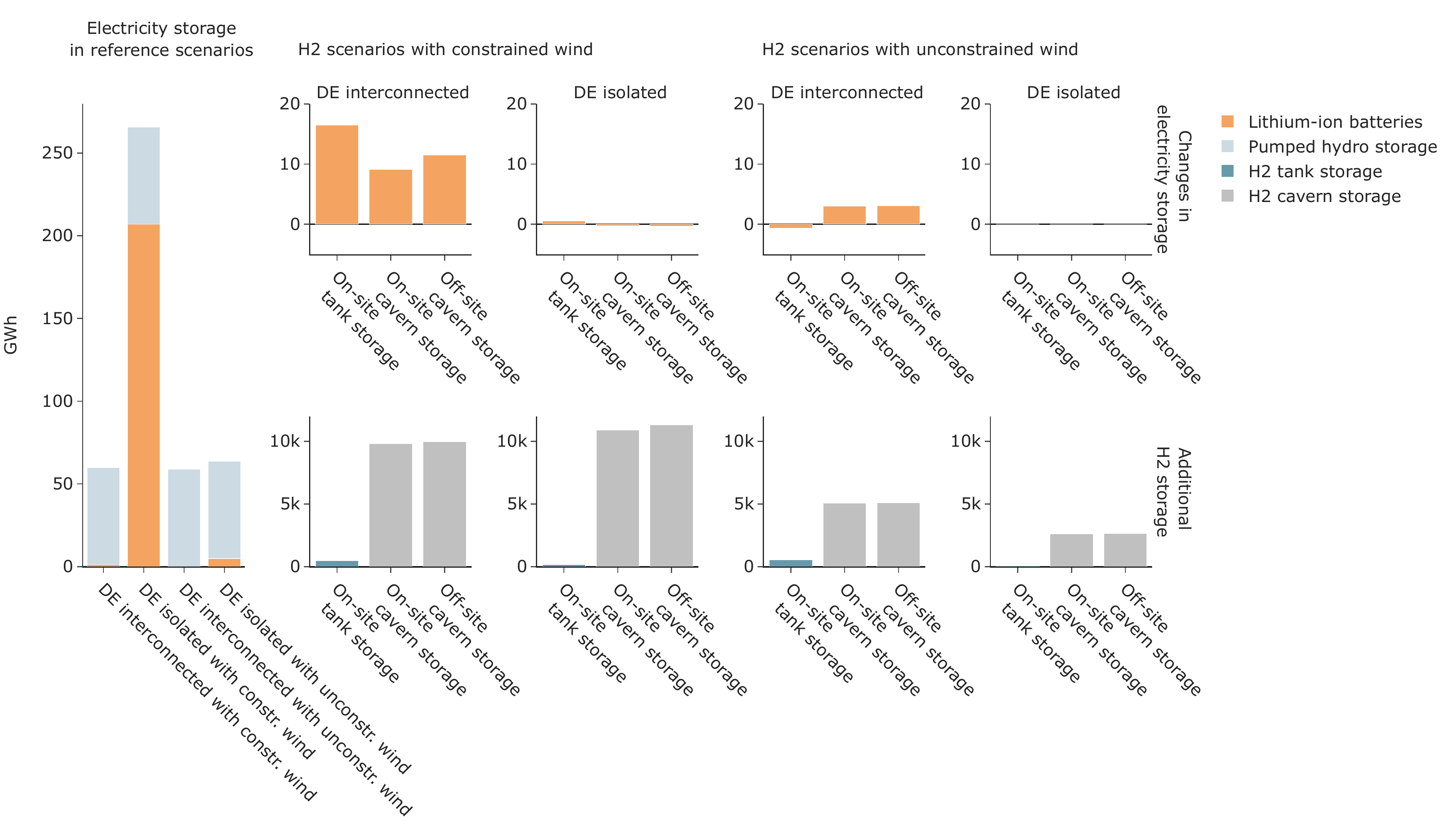}
    \includegraphics[width=\linewidth]{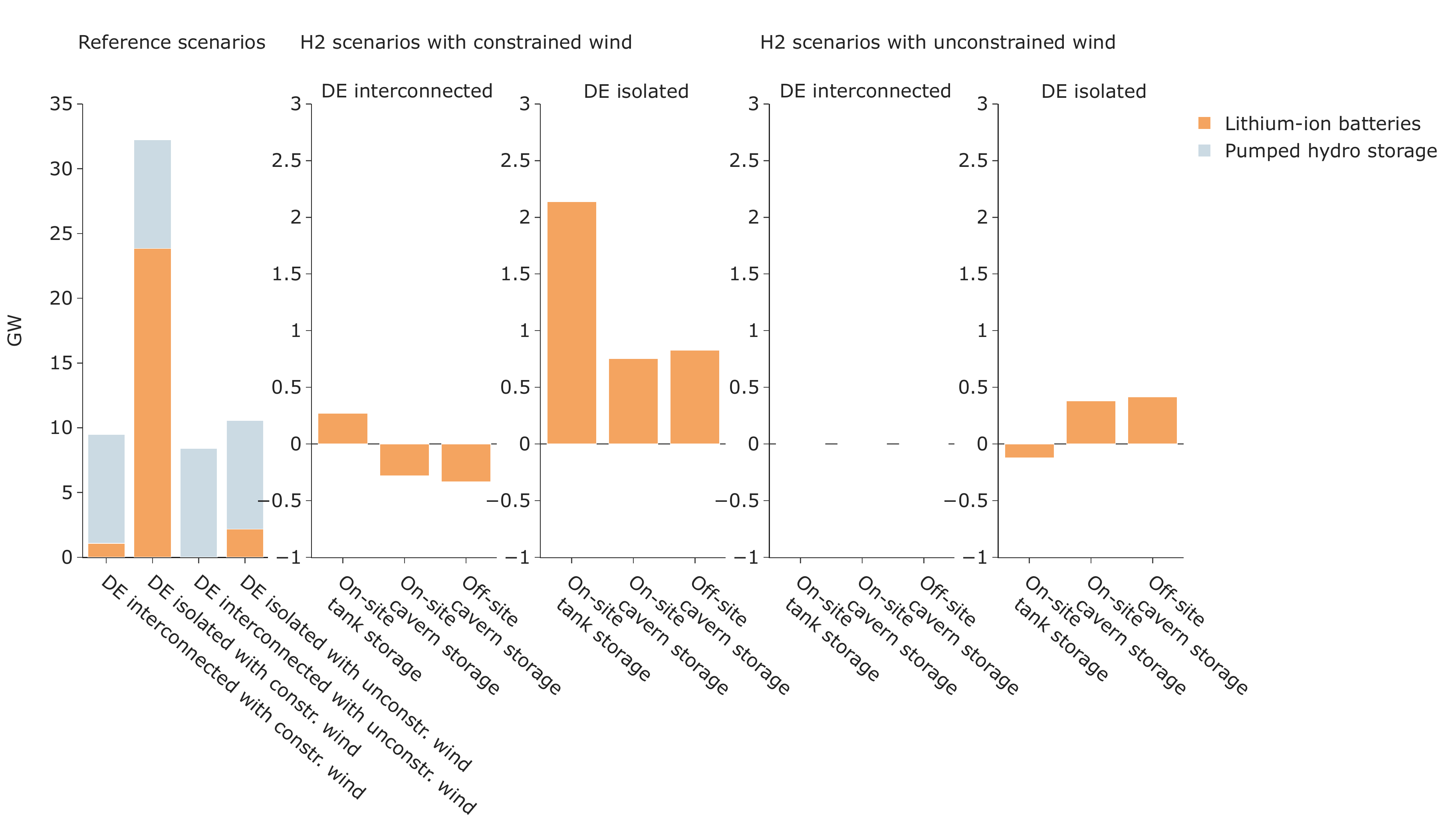}
    \caption{Electricity and hydrogen storage capacities in reference scenarios and changes induced by green hydrogen. Upper panels: storage energy capacity in GWh; lower panels: storage discharging capacity in GW}
    \label{fig:n_sto}
\end{figure}

Figure~\ref{fig:n_sto} shows optimal energy storage capacities and the changes induced by green hydrogen. The energy capacity in GWh in the reference scenarios and the respective changes in electricity storage and hydrogen storage capacities are given in the upper panel, while the lower panel depicts the installed electricity storage discharging capacity in the reference scenarios as well as the hydrogen-induced changes in GW. In the reference scenarios, not much electricity storage is built beyond the exogenously specified pumped hydro storage capacity of 8.4 GW and 58.8~GWh. Pumped hydro storage potential for Germany is assumed to be exhausted, thus no further capacity expansion is considered here. Only in the case where Germany is modeled in isolation and wind power expansion is constrained, there is a sizeable investment in these stationary lithium-ion batteries (23~GW and 206~GWh). In this case, they are used to balance diurnal fluctuations of solar PV.\\

Green hydrogen production slightly increases the optimal lithium-ion energy capacity in the cases where Germany is interconnected with its neighbors, but not when it is modeled as an electric island. This is because electrolyzers can make use of renewable surplus energy to a large extent in the latter case even without additional electricity storage. In contrast, the optimal power rating of lithium-ion storage hardly changes after green hydrogen is added in the interconnected cases, but slightly increases in the 'DE isolated' case where wind power expansion is constrained. Here, additional power generation capacity from battery storage is used to supply electrolyzers with stored solar electricity. Yet, the overall effects of green hydrogen on stationary lithium-ion storage capacities are relatively minor in all cases modeled here. So we find that green hydrogen production neither increases nor decreases the need for electricity storage substantially. This could change if investments in pumped hydro storage were endogenous, since the capacity is already slightly oversized in the TYNDP 2020. If all electricity storage investments were fully endogenous, the hydrogen storage investment options would probably have a larger effect on electricity storage investments.\\ 

Hydrogen storage investments are around two to three orders of magnitude higher than investments in electricity storage, with much higher investments in low-cost caverns than such in tank storage. Hydrogen storage capacities are further higher in scenarios with constrained wind power expansion, as compared to scenarios with unconstrained wind power. This is because wind constraints lead to PV-heavier systems with a larger need for a seasonal balancing of PV surplus energy between summer and winter.

\subsection{Full-load hours of electrolyzers}\label{sec: appendix - flh}

\begin{figure}[H]
    \centering
    \includegraphics[width=\linewidth]{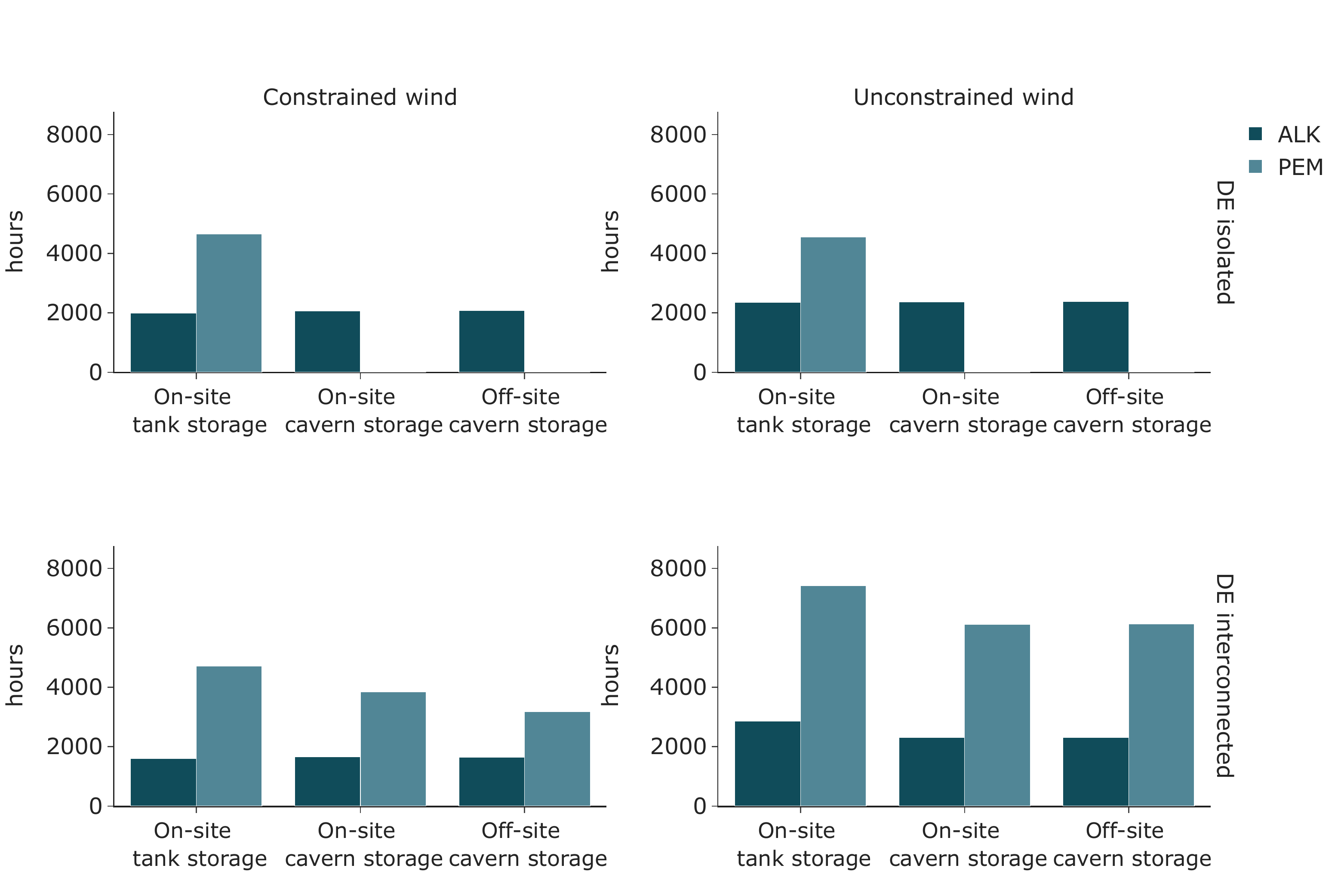}
    \caption{Full-load hours of electrolyzers}
    \label{fig:flh}
\end{figure}

On average, full-load hours of electrolyzers are lower if they are combined with cavern storage (Figure~\ref{fig:flh}), as installed capacities are higher (see Figure~\ref{fig:electrolyzer_cap}). The less energy-efficient alkaline electrolysis technology (ALK) has higher capacity investments, but only 1600 to 2860 full-load hours across all scenarios, since it recovers its lower capital and operational costs even at low full-load hours. In contrast, the (more energy-efficient) PEM technology has higher full-load hours, since it is more expensive to built and therefore has to run more often to recover its costs. In general, PEM electrolyzers have a larger number of full-load hours if installed capacities are bigger (see the two cases 'DE interconnected'). Meanwhile, the increase in alkaline electrolyzer capacity affects its full-load hours only marginally. In general, the results regarding installed electrolyzer capacity and full-load hours strongly depend on the assumed capital and operational costs and efficiencies, as well as the availability of cheap renewable energy. In the investigated settings, the cost differences in electrolyzer technologies outweigh their differences in efficiency - this might change if cost structures change in the future or renewable energy surpluses become scarcer.

 \bibliographystyle{elsarticle-num} 
 \bibliography{manuscript.bib}

\begin{thebibliography}{10}
\expandafter\ifx\csname url\endcsname\relax
  \def\url#1{\texttt{#1}}\fi
\expandafter\ifx\csname urlprefix\endcsname\relax\def\urlprefix{URL }\fi
\expandafter\ifx\csname href\endcsname\relax
  \def\href#1#2{#2} \def\path#1{#1}\fi

\bibitem{deconinck_2018}
H.~de~Coninck, A.~Revi, M.~Babiker, P.~Bertoldi, M.~Buckeridge, A.~Cartwright,
  W.~Dong, J.~Ford, S.~Fuss, J.-C. Hourcade, D.~Ley, R.~Mechler, P.~Newman,
  A.~Revokatova, S.~Schultz, L.~Steg, T.~Sugiyama, {S}trengthening and
  {I}mplementing the {G}lobal {R}esponse, in: V.~Masson-Delmotte, P.~Zhai,
  H.-O. P\"{o}rtner, D.~Roberts, J.~Skea, P.~R. Shukla, A.~Pirani,
  W.~Moufouma-Okia, C.~P{\'{e}}an, R.~Pidcock, S.~Connors, J.~B.~R. Matthews,
  Y.~Chen, X.~Zhou, M.~Gomis, E.~Lonnoy, T.~Maycock, M.~Tignor, T.~Waterfield
  (Eds.), Global {Warming} of {1.5$^{\,\circ}${C}}. {An} {IPCC} {S}pecial
  {R}eport on the {I}mpacts of {G}lobal {W}arming of {1.5$^{\,\circ}${C}}
  {A}bove {P}re-{I}ndustrial {L}evels and {R}elated {G}lobal {G}reenhouse {G}as
  {E}mission {P}athways, in the {C}ontext of {S}trengthening the {G}lobal
  {R}esponse to the {T}hreat of {C}limate {C}hange, {S}ustainable
  {D}evelopment, and {E}fforts to {E}radicate {P}overty, 2018, available at:
  \url{https://www.ipcc.ch/site/assets/uploads/sites/2/2019/05/SR15_Chapter4_High_Res.pdf}.

\bibitem{IPCC_2022}
P.~Shukla, J.~Skea, R.~Slade, A.~Al~Khourdajie, R.~van Diemen, D.~McCollum,
  M.~Pathak, S.~Some, P.~Vyas, R.~Fradera, M.~Belkacemi, A.~Hasija, G.~Lisboa,
  S.~Luz, J.~e. Malley, Contribution of {W}orking {G}roup {III} to the {S}ixth
  {A}ssessment {R}eport of the {I}ntergovernmental {P}anel on {C}limate
  {C}hange, in: IPCC, 2022: Climate Change 2022: Mitigation of Climate Change.,
  Cambridge University Press, Cambridge, UK and New York, NY, USA, 2022,
  \url{doi: 10.1017/9781009157926}.

\bibitem{schlund_2022}
D.~Schlund, P.~Theile, Simultaneity of green energy and hydrogen production:
  Analysing the dispatch of a grid-connected electrolyser, Energy Policy 166 (7
  2022).
\newblock \href {https://doi.org/10.1016/j.enpol.2022.113008}
  {\path{doi:10.1016/j.enpol.2022.113008}}.

\bibitem{welder_2018}
L.~Welder, D.~S. Ryberg, L.~Kotzur, T.~Grube, M.~Robinius, D.~Stolten,
  {S}patio-{T}emporal {O}ptimization of a {F}uture {E}nergy {S}ystem for
  {P}ower-to-{H}ydrogen {A}pplications in {G}ermany, Energy 158 (2018)
  1130--1149.
\newblock \href {https://doi.org/10.1016/j.energy.2018.05.059}
  {\path{doi:10.1016/j.energy.2018.05.059}}.

\bibitem{samsatli_2016}
S.~Samsatli, I.~Staffell, N.~J. Samsatli, {O}ptimal {D}esign and {O}peration of
  {I}ntegrated {W}ind-{H}ydrogen-{E}lectricity {N}etworks for {D}ecarbonising
  the {D}omestic {T}ransport {S}ector in {G}reat {B}ritain, International
  Journal of Hydrogen Energy 41~(1) (2016) 447--475.
\newblock \href {https://doi.org/10.1016/j.ijhydene.2015.10.032}
  {\path{doi:10.1016/j.ijhydene.2015.10.032}}.

\bibitem{robinius_2017}
M.~Robinius, A.~Otto, K.~Syranidis, D.~S. Ryberg, P.~Heuser, L.~Welder,
  T.~Grube, P.~Markewitz, V.~Tietze, D.~Stolten, {L}inking the {P}ower and
  {T}ransport {S}ectors - {P}art 2: {M}odelling a {S}ector {C}oupling
  {S}cenario for {G}ermany, Energies 10~(7) (2017) 957.
\newblock \href {https://doi.org/10.3390/en10070957}
  {\path{doi:10.3390/en10070957}}.

\bibitem{emonts_2019}
B.~Emonts, M.~Reu{\ss}, P.~Stenzel, L.~Welder, F.~Knicker, T.~Grube,
  K.~G{\"{o}}rner, M.~Robinius, D.~Stolten, {F}lexible {S}ector {C}oupling with
  {H}ydrogen: {A} {C}limate-{F}riendly {F}uel {S}upply for {R}oad {T}ransport,
  International Journal of Hydrogen Energy 44~(26) (2019) 12918--12930.
\newblock \href {https://doi.org/10.1016/j.ijhydene.2019.03.183}
  {\path{doi:10.1016/j.ijhydene.2019.03.183}}.

\bibitem{glenk_2019}
G.~Glenk, S.~Reichelstein, {E}conomics of {C}onverting {R}enewable {P}ower to
  {H}ydrogen, Nature Energy 4~(3) (2019) 216--222.
\newblock \href {https://doi.org/10.1038/s41560-019-0326-1}
  {\path{doi:10.1038/s41560-019-0326-1}}.

\bibitem{yang_2007}
C.~Yang, J.~Ogden, {D}etermining the {L}owest-{C}ost {H}ydrogen {D}elivery
  {M}ode, International Journal of Hydrogen Energy 32~(2) (2007) 268--286.
\newblock \href {https://doi.org/10.1016/j.ijhydene.2006.05.009}
  {\path{doi:10.1016/j.ijhydene.2006.05.009}}.

\bibitem{reuss_2017}
M.~Reu{\ss}, T.~Grube, M.~Robinius, P.~Preuster, P.~Wasserscheid, D.~Stolten,
  {S}easonal {S}torage and {A}lternative {C}arriers: {A} {F}lexible {H}ydrogen
  {S}upply {C}hain {M}odel, Applied Energy 200 (2017) 290--302.
\newblock \href {https://doi.org/10.1016/j.apenergy.2017.05.050}
  {\path{doi:10.1016/j.apenergy.2017.05.050}}.

\bibitem{sens_2022}
L.~Sens, Y.~Piguel, U.~Neuling, S.~Timmerberg, K.~Wilbrand, M.~Kaltschmitt,
  Cost minimized hydrogen from solar and wind – production and supply in the
  {European} catchment area, Energy Conversion and Management 265 (2022)
  115742.
\newblock \href {https://doi.org/10.1016/j.enconman.2022.115742}
  {\path{doi:10.1016/j.enconman.2022.115742}}.

\bibitem{Zhang_2020}
C.~Zhang, J.~B. Greenblatt, W.~Wei, J.~Eichman, S.~Saxena, M.~Muratori, O.~J.
  Guerra, {F}lexible {G}rid-{B}ased {E}lectrolysis {H}ydrogen {P}roduction for
  {F}uel {C}ell {V}ehicles {R}educes {C}osts and {G}reenhouse {G}as
  {E}missions, Applied Energy 278 (2020) 115651.
\newblock \href {https://doi.org/10.1016/j.apenergy.2020.115651}
  {\path{doi:10.1016/j.apenergy.2020.115651}}.

\bibitem{runge_2019}
P.~Runge, C.~S\"{o}lch, J.~Albert, P.~Wasserscheid, G.~Z\"{o}ttl, V.~Grimm,
  {E}conomic {C}omparison of {D}ifferent {E}lectric {F}uels for {E}nergy
  {S}cenarios in 2035, Applied Energy 233-234 (2019) 1078--1093.
\newblock \href {https://doi.org/10.1016/j.apenergy.2018.10.023}
  {\path{doi:10.1016/j.apenergy.2018.10.023}}.

\bibitem{michalski_2017}
J.~Michalski, U.~B{\"{u}}nger, F.~Crotogino, S.~Donadei, G.-S. Schneider,
  T.~Pregger, K.-K. Cao, D.~Heide, {H}ydrogen {G}eneration by {E}ectrolysis and
  {S}torage in {S}alt {C}averns: {P}otentials, {E}conomics and {S}ystems
  {A}spects with {R}egard to the {G}erman {E}nergy {T}ransition, International
  Journal of Hydrogen Energy 42~(19) (2017) 13427--13443.
\newblock \href {https://doi.org/10.1016/j.ijhydene.2017.02.102}
  {\path{doi:10.1016/j.ijhydene.2017.02.102}}.

\bibitem{vomscheidt_2022}
F.~{vom Scheidt}, J.~Qu, P.~Staudt, D.~S. Mallapragada, C.~Weinhardt,
  Integrating hydrogen in single-price electricity systems: The effects of
  spatial economic signals, Energy Policy 161 (2022) 112727.
\newblock \href {https://doi.org/doi.org/10.1016/j.enpol.2021.112727}
  {\path{doi:doi.org/10.1016/j.enpol.2021.112727}}.

\bibitem{weimann_2021}
L.~Weimann, P.~Gabrielli, A.~Boldrini, G.~J. Kramer, M.~Gazzani, Optimal
  hydrogen production in a wind-dominated zero-emission energy system, Advances
  in Applied Energy 3 (2021) 100032.

\bibitem{breder_2022}
M.~S. Breder, F.~Meurer, M.~Bucksteeg, C.~Weber,
  \href{https://ideas.repec.org/p/dui/wpaper/2203.html}{{Spatial Incentives for
  Power-to-hydrogen through Market Splitting}}, EWL Working Paper 2203,
  University of Duisburg-Essen, Chair for Management Science and Energy
  Economics (Jul. 2022).
\newline\urlprefix\url{https://ideas.repec.org/p/dui/wpaper/2203.html}

\bibitem{lieberwirth_2023}
M.~Lieberwirth, H.~Hobbie,
  \href{https://www.sciencedirect.com/science/article/pii/S0959652623009150}{Decarbonizing
  the industry sector and its effect on electricity transmission grid
  operation—implications from a model based analysis for {G}ermany}, Journal
  of Cleaner Production 402 (2023) 136757.
\newblock \href {https://doi.org/https://doi.org/10.1016/j.jclepro.2023.136757}
  {\path{doi:https://doi.org/10.1016/j.jclepro.2023.136757}}.
\newline\urlprefix\url{https://www.sciencedirect.com/science/article/pii/S0959652623009150}

\bibitem{brown_2018}
T.~Brown, D.~Schlachtberger, A.~Kies, S.~Schramm, M.~Greiner, Synergies of
  sector coupling and transmission reinforcement in a cost-optimised, highly
  renewable {European} energy system, Energy 160 (2018) 720--739.
\newblock \href {https://doi.org/10.1016/j.energy.2018.06.222}
  {\path{doi:10.1016/j.energy.2018.06.222}}.

\bibitem{durakovic_2023}
G.~Durakovic, P.~C. del Granado, A.~Tomasgard, Powering europe with north sea
  offshore wind: The impact of hydrogen investments on grid infrastructure and
  power prices, Energy 263 (2023) 125654.

\bibitem{stoeckl_2021}
F.~Stöckl, W.-P. Schill, A.~Zerrahn, Optimal supply chains and power sector
  benefits of green hydrogen, Scientific Reports 11~(1) (2021) 14191.
\newblock \href {https://doi.org/10.1038/s41598-021-92511-6}
  {\path{doi:10.1038/s41598-021-92511-6}}.

\bibitem{shirizadeh_2022}
B.~Shirizadeh, P.~Quirion, The importance of renewable gas in achieving
  carbon-neutrality: Insights from an energy system optimization model, Energy
  255 (2022) 124503.
\newblock \href {https://doi.org/10.1016/j.energy.2022.124503}
  {\path{doi:10.1016/j.energy.2022.124503}}.

\bibitem{peterssen_2022}
F.~Peterssen, M.~Schlemminger, C.~Lohr, R.~Niepelt, A.~Bensmann,
  R.~Hanke-Rauschenbach, R.~Brendel, Hydrogen supply scenarios for a climate
  neutral energy system in {G}ermany, International Journal of Hydrogen Energy
  47~(28) (2022) 13515--13523.

\bibitem{luderer_2022}
G.~Luderer, S.~Madeddu, L.~Merfort, F.~Ueckerdt, M.~Pehl, R.~Pietzcker,
  M.~Rottoli, F.~Schreyer, N.~Bauer, L.~Baumstark, C.~Bertram, A.~Dirnaichner,
  F.~Humpenöder, A.~Levesque, A.~Popp, R.~Rodrigues, J.~Strefler, E.~Kriegler,
  Impact of declining renewable energy costs on electrification in low-emission
  scenarios, Nature Energy 7~(1) (2022) 32--42.
\newblock \href {https://doi.org/10.1038/s41560-021-00937-z}
  {\path{doi:10.1038/s41560-021-00937-z}}.

\bibitem{zerrahn_2017}
A.~Zerrahn, W.-P. Schill, Long-run power storage requirements for high shares
  of renewables: review and a new model, Renewable and Sustainable Energy
  Reviews 79 (2017) 1518 -- 1534.
\newblock \href {https://doi.org/10.1016/j.rser.2016.11.098}
  {\path{doi:10.1016/j.rser.2016.11.098}}.

\bibitem{gaete_2021}
C.~Gaete-Morales, M.~Kittel, A.~Roth, W.-P. Schill, {DIETERpy}: a {Python}
  framework for the {Dispatch} and {Investment} {Evaluation} {Tool} with
  {Endogenous} {Renewables}, SoftwareX 15 (2021) 100784.
\newblock \href {https://doi.org/10.1016/j.softx.2021.100784}
  {\path{doi:10.1016/j.softx.2021.100784}}.

\bibitem{schill_2018}
W.-P. Schill, A.~Zerrahn, Long-run power storage requirements for high shares
  of renewables: Results and sensitivities, Renewable and Sustainable Energy
  Reviews 83 (2018) 156 -- 171.
\newblock \href {https://doi.org/10.1016/j.rser.2017.05.205}
  {\path{doi:10.1016/j.rser.2017.05.205}}.

\bibitem{schill_heating_2020}
W.-P. Schill, A.~Zerrahn, Flexible electricity use for heating in markets with
  renewable energy, Applied Energy 266 (2020) 114571.
\newblock \href {https://doi.org/10.1016/j.apenergy.2020.114571}
  {\path{doi:10.1016/j.apenergy.2020.114571}}.

\bibitem{say_2020}
K.~Say, W.-P. Schill, M.~John, Degrees of displacement: The impact of household
  {PV} battery prosumage on utility generation and energy storage, Applied
  Energy 276 (2020) 115466.
\newblock \href {https://doi.org/10.1016/j.apenergy.2020.115466}
  {\path{doi:10.1016/j.apenergy.2020.115466}}.

\bibitem{BMWK_2022}
{BMWK}, {Überblickspapier Osterpaket}, {German Federal Ministry for Economic
  Affairs and Climate Action,
  \url{https://www.bmwk.de/Redaktion/DE/Downloads/Energie/0406_ueberblickspapier_osterpaket.html}}
  ({A}pril 2022).

\bibitem{kittel_2022}
M.~Kittel, W.-P. Schill, Renewable energy targets and unintended storage
  cycling: Implications for energy modeling, Iscience 25~(4) (2022) 104002.

\bibitem{entsoe_2022}
ENTSO-E, ENTSO-G, {TYNDP} 2020 – {S}cenarios [online],
  \url{https://www.entsos-tyndp2020-scenarios.eu/} (accessed: 27.07.2022).

\bibitem{pietzcker_2021}
R.~C. Pietzcker, J.~Feuerhahn, L.~Haywood, B.~Knopf, F.~Leukhardt, G.~Luderer,
  S.~Osorio, M.~Pahle, R.~Dias Bleasby~Rodrigues, O.~Edenhofer, Notwendige
  co2-preise zum erreichen des europ{\"a}ischen klimaziels 2030 (2021).

\bibitem{opendata}
{OPSD}, {Open Power System Data},
  \url{https://data.open-power-system-data.org/} (accessed: 27.07.2022).

\bibitem{wiese_2019}
F.~Wiese, I.~Schlecht, W.-D. Bunke, C.~Gerbaulet, L.~Hirth, M.~Jahn, F.~Kunz,
  C.~Lorenz, J.~Mühlenpfordt, J.~Reimann, W.-P. Schill, Open power system data
  – frictionless data for electricity system modelling, Applied Energy 236
  (2019) 401--409.
\newblock \href {https://doi.org/10.1016/j.apenergy.2018.11.097}
  {\path{doi:10.1016/j.apenergy.2018.11.097}}.

\bibitem{poestges_2022}
A.~Pöstges, M.~Bucksteeg, O.~Ruhnau, D.~Böttger, M.~Haller, E.~Künle,
  D.~Ritter, R.~Schmitz, M.~Wiedmann, Phasing out coal: An impact analysis
  comparing five large-scale electricity market models, Applied Energy 319
  (2022) 119215.
\newblock \href {https://doi.org/10.1016/j.apenergy.2022.119215}
  {\path{doi:10.1016/j.apenergy.2022.119215}}.

\bibitem{saba_2018}
S.~M. Saba, M.~M{\"u}ller, M.~Robinius, D.~Stolten, The investment costs of
  electrolysis--a comparison of cost studies from the past 30 years,
  International journal of hydrogen energy 43~(3) (2018) 1209--1223.

\bibitem{schlachtberger_2017}
D.~Schlachtberger, T.~Brown, S.~Schramm, M.~Greiner, The benefits of
  cooperation in a highly renewable {European} electricity network, Energy 134
  (2017) 469--481.
\newblock \href {https://doi.org/10.1016/j.energy.2017.06.004}
  {\path{doi:10.1016/j.energy.2017.06.004}}.

\bibitem{roth_2023}
A.~Roth, W.-P. Schill, Geographical balancing of wind power decreases storage
  needs in a 100\% renewable {European} power sector, iScience (2023)
  107074\href {https://doi.org/10.1016/j.isci.2023.107074}
  {\path{doi:10.1016/j.isci.2023.107074}}.

\bibitem{schill_2020}
W.-P. Schill, Electricity storage and the renewable energy transition, Joule
  4~(10) (2020) 2059--2064.
\newblock \href {https://doi.org/10.1016/j.joule.2020.07.022}
  {\path{doi:10.1016/j.joule.2020.07.022}}.

\bibitem{brown_2021}
T.~Brown, L.~Reichenberg, Decreasing market value of variable renewables can be
  avoided by policy action, Energy Economics 100 (2021) 105354.
\newblock \href {https://doi.org/10.1016/j.eneco.2021.105354}
  {\path{doi:10.1016/j.eneco.2021.105354}}.

\bibitem{lopez_prol_2021}
J.~L\'{o}pez~Prol, W.-P. Schill, The economics of variable renewable energy and
  electricity storage, Annual Review of Resource Economics 13~(1) (2021)
  443--467.
\newblock \href {https://doi.org/10.1146/annurev-resource-101620-081246}
  {\path{doi:10.1146/annurev-resource-101620-081246}}.

\bibitem{gils_2021}
H.~C. Gils, H.~Gardian, J.~Schmugge, Interaction of hydrogen infrastructures
  with other sector coupling options towards a zero-emission energy system in
  {Germany}, Renewable Energy 180 (2021) 140--156.
\newblock \href {https://doi.org/10.1016/j.renene.2021.08.016}
  {\path{doi:10.1016/j.renene.2021.08.016}}.

\bibitem{husarek_2021}
D.~Husarek, J.~Schmugge, S.~Niessen, Hydrogen supply chain scenarios for the
  decarbonisation of a german multi-modal energy system, International Journal
  of Hydrogen Energy 46 (2021) 38008--38025.
\newblock \href {https://doi.org/10.1016/j.ijhydene.2021.09.041}
  {\path{doi:10.1016/j.ijhydene.2021.09.041}}.

\bibitem{brandle_2021}
G.~Brändle, M.~Schönfisch, S.~Schulte, Estimating long-term global supply
  costs for low-carbon hydrogen, Applied Energy 302 (11 2021).
\newblock \href {https://doi.org/10.1016/j.apenergy.2021.117481}
  {\path{doi:10.1016/j.apenergy.2021.117481}}.

\bibitem{BMWI_2020}
{BMWi}, {The National Hydrogen Strategy}, {German Federal Ministry for Economic
  Affairs and Energy,
  \url{https://www.bmwk.de/Redaktion/EN/Publikationen/Energie/the-national-hydrogen-strategy.html}}
  ({J}une 2020).

\bibitem{odenweller_2022}
A.~Odenweller, F.~Ueckerdt, G.~F. Nemet, M.~Jensterle, G.~Luderer,
  Probabilistic feasibility space of scaling up green hydrogen supply, Nature
  Energy 7~(9) (2022) 854--865.

\bibitem{reuss_2019}
M.~Reuß, T.~Grube, M.~Robinius, D.~Stolten, A hydrogen supply chain with
  spatial resolution: Comparative analysis of infrastructure technologies in
  {Germany}, Applied Energy 247 (2019) 438--453.
\newblock \href {https://doi.org/10.1016/j.apenergy.2019.04.064}
  {\path{doi:10.1016/j.apenergy.2019.04.064}}.

\bibitem{brey_2021}
J.~Brey, Use of hydrogen as a seasonal energy storage system to manage
  renewable power deployment in {S}pain by 2030, International Journal of
  Hydrogen Energy 46~(33) (2021) 17447--17457.

\bibitem{pavic_2022}
I.~Pavi{\'c}, N.~{\v{C}}ovi{\'c}, H.~Pand{\v{z}}i{\'c}, {PV}--battery-hydrogen
  plant: Cutting green hydrogen costs through multi-market positioning, Applied
  Energy 328 (2022) 120103.

\end{thebibliography}

\end{document}